\RequirePackage{pdf14}
\documentclass[11pt,a4paper]{article}
\usepackage[a4paper]{geometry}
\usepackage{setspace}
\usepackage{array}
\usepackage[centertags,reqno]{amsmath}
\usepackage{graphics,subfigure}
\usepackage{colortbl}
\usepackage{hyperref}
\usepackage{appendix}
\usepackage{threeparttable}
\usepackage{rotating}
\usepackage{dcolumn}
\usepackage{natbib}
\usepackage{hyperref}
\usepackage{longtable}
\usepackage{caption}
\usepackage{pdflscape}
\usepackage{xcolor}
\usepackage{authblk}
\usepackage{amscd,latexsym,amsfonts}
\usepackage{verbatim,color}
\usepackage{booktabs}
\usepackage{epsfig}
\usepackage{epstopdf}
\usepackage{adjustbox}
\usepackage{multirow}
\usepackage{float,lscape}
\usepackage{bigstrut}
\usepackage{array}
\usepackage{tabularx}
\usepackage{fancyhdr}
\usepackage{lipsum}
\usepackage{bm}
\usepackage{amssymb}
\usepackage{ragged2e}
\usepackage{pgfplots}
\pgfplotsset{compat=newest,xlabel near ticks,ylabel near ticks,width=15cm}
\usepackage{tikz}
\usetikzlibrary{plotmarks}
\usepackage{etex}
\usepackage{filecontents}
\usetikzlibrary{calc}
\usepackage{environ}
\usepackage{xstring}

\usepackage{xcolor}
\usepackage{bchart}

\newcolumntype{d}[1]{D{.}{.}{#1}}
\newcolumntype{p}[1]{D{(}{(}{-1}}

\newcolumntype{L}[1]{>{\raggedright\let\newline\\\arraybackslash\hspace{0pt}}m{#1}}
\newcolumntype{C}[1]{>{\centering\let\newline\\\arraybackslash\hspace{0pt}}m{#1}}
\newcolumntype{R}[1]{>{\raggedleft\let\newline\\\arraybackslash\hspace{0pt}}m{#1}}
\begin{document}
	\title{Bridging the Covid-19 Data and the Epidemiological Model using Time-Varying Parameter SIRD Model   
		\\ 
		\vspace{0.5cm}
	%	{\normalsize Revise and Resubmit Requested from Journal of Econometrics}
	}
	
	\author[,1]{ Cem \c{C}akmakl{\i} \thanks{
			Correspondence to: Cem \c{C}akmakl{\i}, Ko\c{c} University, Rumelifeneri Yolu 34450 Sarıyer Istanbul Turkey, 
			e--mail: ccakmakli@ku.edu.tr. }
	}
	\author[,2]{Yasin \c{S}im\c{s}ek \thanks{e--mail: yasin.simsek@duke.edu}}
	\affil[1]{\emph{Ko\c{c} University}}
	\affil[2]{\emph{Duke University}}

	\maketitle
	\justifying
	\begin{abstract} \noindent
This paper extends the canonical model of epidemiology, the SIRD model, to allow for time-varying parameters for real-time measurement and prediction of the trajectory of the Covid-19 pandemic. Time variation in model parameters is captured using the generalized autoregressive score modeling structure designed for the typical daily count data related to the pandemic. The resulting specification permits a flexible yet parsimonious model with a low computational cost. The model is extended to allow for unreported cases using a mixed-frequency setting. Results suggest that these cases' effects on the parameter estimates might be sizeable. Full sample results show that the flexible framework accurately captures the successive waves of the pandemic. A real-time exercise indicates that the proposed structure delivers timely and precise information on the pandemic's current stance. This superior performance, in turn, transforms into accurate predictions of the confirmed and death cases.
		\end{abstract}

	\newpage
	
	\setcounter{page}{1}
%	\onehalfspacing
%\linespread{1.25}
\setstretch{1.5}
	
\section{Introduction}
The outbreak of the new coronavirus, the Covid-19 pandemic, is one of the most severe health crises the world has encountered in the last decades. Since the onset of the pandemic in early January 2020, it has exhibited a varying pattern for many reasons. First, countries have repeatedly taken various measures to reduce the transmission of the virus. These measures involve complete lockdowns, such as full closure of business and curfew, and partial lockdown that implies a partial closure of daily routines. These interventions seem to mitigate the spread of the virus at various stages of the pandemic to the extent that people comply with the 'shelter-in-place' orders. Second, mutations of the virus might lead to changes in its main characteristics. Furthermore, the death rate seems to be partially lowered thanks to vaccination campaigns, increasing medical knowledge, and ongoing research on the virus.

While the pandemic evolves rapidly with successive waves of infections, efficient and timely monitoring is crucial. Making prompt and effective decisions of imposing or relaxing lockdown measures for policymakers and taking timely precautions for individuals critically relies on knowledge about the pandemic. Therefore, epidemiological models for estimating and, perhaps even more crucially, for predicting the pandemic's trajectory come to the forefront. Conventional statistical epidemiology models mainly involve structural parameters that remain constant throughout the pandemic. However, suppose these interventions are effective and cause changes in the pandemic's natural course. Essentially, even if the underlying structural parameters related to the pandemic remain unaltered, there might be various reasons for the resulting estimated parameter to be time-varying. These include changes in the people's attitude towards the disease\footnote{\cite{arias2021bayesian} uses the terminology of behavioral parameters to refer to the behavioral aspects that lead to time variation in the structural parameters. Similarly, \cite{avery2020economist} refers to the changes in the parameters as potential endogeneity of these parameters as the precautions taken could be a function of current cases. We thoroughly discuss these issues and the inability to measure the observations of units/compartments, such as the number of susceptible people required for estimating the models in Section~\ref{subsec:Motivation}.}. Besides, the virus might undergo some mutations which alter the contagiousness and fatality of the virus; see, for example, \cite{karim2021omicron} for the recent Omicron variant of the virus. Hence, these mutations translate into changes in key structural model parameters that alter smoothly with the changing weight of infections of each virus variant. These observations are the departure point of this paper. Specifically, we develop a computationally simple and statistically coherent model that allows for time variation in the epidemiological model parameters. 

We start our analysis by confronting a simple version of the workhorse epidemiological model with the existing data. From the perspective of econometrics, we specify a counting process for modeling the course of the Covid-19 pandemic for the US based on the SIRD model, which is an abbreviation representing the four identified states of the pandemic as Susceptible, Infected, Recovered, and Death. The SIRD model depicts these states' evolution depending on the number of infected individuals; see \cite{kermack1927, Allen-2008}. Provided that these daily counts of susceptible, infected, recovered, and death cases are available, the model is well-identified conditional on the infected cases' initial value. The pandemic's course is determined by the contestation of these forces, i.e., the parameters governing infection and resolution (either recovery or death) rates. As a result, if the rate of infection (multiplied by the share of susceptible people in the population) is larger than the resolution rate, the number of infections evolves according to a nonstationary process representing the virus's increasing spread. In contrast, the opposite case results in a stationary process. Therefore, we opt for a Bayesian estimation strategy of the parameters for computing reliable credible intervals for inference conditional on the available data rather than utilizing asymptotic analysis.    

Equipped with these tools, we extend the econometric model by allowing for time variation in the structural parameters by resorting to the Generalized Autoregressive Score (henceforth GAS) modeling framework, which is a class of observation-driven models. The proposed model permits a flexible yet feasible framework to track structural parameters' evolution timely and accurate. A relevant aspect of our specification is its relatively low computational cost. The computational cost might be crucial, especially at the beginning of the pandemic, when the data is scarce, plaguing the inference of flexible models, and the uncertainty is overwhelming. Furthermore, since the typical Covid-19 dataset exhibits a sizeable seasonal pattern, the model is extended to capture these stark seasonal variations using a seasonal component for each parameter by resorting to the frequency domain and GAS framework model structures.
We extend the model further, taking undocumented cases (as these infected individuals do not show symptoms) into consideration by exploiting the information on testing for infection following \cite{Grewelle-2020} and on the number of excess deaths, i.e., the total number of deaths caused by the pandemic that is in excess of the reported death cases. Since the typical excess death data is at the weekly frequency, unlike the remaining daily counts, the resulting extension permits a mixed frequency time-varying parameter epidemiological model blending datasets with mixed frequency. Finally, we also consider the model in a multi-country setting. In particular, we put forward the factor TVP-SIRD model, where we consider four countries jointly, focusing on the common and idiosyncratic patterns of the model parameters. The resulting model can efficiently capture the common behavior in the diffusion of the pandemic throughout the world for monitoring the stance of the pandemic from a global perspective. 

We consider the US dataset related to the pandemic to demonstrate the proposed framework's efficacy. The US has ample experience in containing the virus with differential momentum in fighting the disease at various pandemic stages. While the initial US response to the pandemic in terms of imposing restrictions and establishing a containment infrastructure such as mass testing was not fast, the US was among the first countries to start a massive inoculation campaign at the start of 2021. These distinct experiences provide a testing ground with various patterns to examine the proposed model's efficacy. Our results indicate that the model parameters exhibit substantial changes over time. The virus' transmission had reduced considerably with the start of 2021 thanks to the massive vaccination campaign in the US. However, this pattern is interrupted by the two successive waves of the Delta variant and then the Omicron variant, with a rapid increase in virus transmission and a relatively low death rate captured by our model. The mixed frequency extension of the model that also captures unreported cases suggests that the actual number of infected cases is potentially as high as three times more than the reported cases for some periods. Provided by the wide 95\% credibility set ranging from two to five, this finding is consistent with the estimates around four provided by the Center for Diseases Control and Prevention (CDC) for the period until the end of 2020, see \cite{Reese_etal_2020}. The multi-country analysis involving Germany, Italy, and Brazil on top of the US using the factor TVP-SIRD model indicates that the pandemic diffusion shares a sizable common pattern with European countries, including Germany and Italy. However, the diffusion of the pandemic is relatively more idiosyncratic in Brazil.

We examine the model's performance in real-time by conducting a recursive estimation and forecasting exercise with real-time datasets predicting 1- to 30-day ahead number of confirmed and death cases. Results indicate that the proposed model with time-varying parameters provides timely information on the pandemic's current stance ahead of the competing models. While the results are relatively mixed for long horizons from  2-week up to a 1-month ahead, our model yields superior forecasting performance up to 2-week horizon against many competitors, including a linear Gaussian state-space model and a subclass of our model framework. Therefore, the resulting model is instrumental in providing crucial information on the stance of the weeks ahead of the pandemic. We further confront the weekly predictions using our model with those of the models that are included in the Forecast Hub\footnote{\url{https://covid19forecasthub.org/}}, which is a forecast data repository with the predictions created by dozens of leading infectious disease modeling teams from around the globe, in coordination with the CDC. Our comparison indicates that the proposed TVP-SIRD model outperforms more than half of those leading models when 1-week ahead forecasts are considered. However, this outperformance reduces gradually with the increasing horizon. The Forecast Hub additionally provides an ensemble model; a forecast combination scheme generated using individual models built on different assumptions. The results show that our model provides superior forecasts, specifically at the onset of the pandemic when the data is scarce, reflecting our proposed framework's flexible yet parsimonious model structure. 

The literature on estimating the SIRD model (with fixed parameters) and variants to evaluate the current stance of the Covid-19 pandemic has exploded since its outbreak. Relatively earlier analyses include \cite{Read-etal-2020} and \cite{lourenco2020fundamental}, which estimate a SIRD-based model with the data from China for the former and from the UK and Italy for the latter using a likelihood-based inference strategy. \cite{wu2020nowcasting} blend data related to Covid-19 for China with mobility data and estimate the epidemiological model using Bayesian inference to predict the spread of the infection domestically and internationally. \cite{li2020substantial} conduct a similar analysis employing a modified SIRD model together with a network structure and mobility data to uncover the size of the undocumented cases; see also \cite{hortaccsu2020estimating}. \cite{zhang2020prediction} extend the standard SIRD model with many additional compartments and estimate some parameters using Bayesian inference. Identification of the model parameters in these models hinges upon the data availability for each compartment. Otherwise, parameter values are set based on the pandemic's stylized facts; see \cite{manski2020estimating, Atkeson-2020, KOROLEV2020}.

Several factors might lead to the time variation in the parameters of the epidemiological model. On the one hand, lockdown measures implemented by the policymakers isolate the infected from the susceptible individuals. Therefore, the parameter governing the infection rate, that is, the average number of contacts of an individual, is likely to alter with lockdown conduct, see \cite{hale2020variation} for example. On the other hand, advancements in the fight against Covid-19, including the recovery of drugs and vaccination, could effectively mitigate the course of the disease. In addition, the installment or the lack of medical equipment such as ventilators might alter the rate of recovery or, in other words, the duration of the state of being infected, see for example \cite{Greenhalgh-Day-2017} on time variation in recovery rates. Accordingly, \cite{anastassopoulou2020data} use a least-squares-based approach on a rolling window of daily observations. They document the time variation of parameters in the SIRD-based model using Chinese data. \cite{tan2020} also employ a similar but more articulated rolling window strategy to capture the time variation in the model parameters. Other frameworks with time-varying model parameters almost exclusively allow for the time variation only in the infection rate. An application before the Covid-19 outbreak includes, for example, \cite{xu2016bayesian} among others, who utilize a Gaussian process prior to the incidence rate involving the infection rate using a Bayesian nonparametric structure. In the context of the Covid-19 pandemic, \cite{Kucharski-etal-2020} estimate a modified SIRD model using a parameter-driven model framework allowing the infection rate to follow a geometric random walk with the remaining parameters kept as constant; see \cite{kucinskas2020tracking} for a similar approach.
Similarly, \cite{yang2020short} and \cite{fernandez2020estimating} allow for time variation in the infection rate, keeping the remaining parameters constant. \cite{arias2021bayesian}, on the other hand, extends the model with time variation in the remaining parameters and provides a Bayesian inference methodology of the resulting model using a particle filter. \cite{LIU-2020} proposes an econometric specification where the growth rate of infections follows an autoregressive process around a deterministic trend with a structural break.

In this paper, we propose an alternative modeling strategy to capture the time variation in the structural parameters of the SIRD model. On the one hand, our modeling framework is statistically consistent with the typical count data structure related to the pandemic, unlike the models that either employ least-squares or likelihood-based inference using Gaussian distribution, that is, the Kalman filter. On the other hand, our framework is computationally inexpensive, unlike the models that are statistically consistent but computationally costly such as the particle filter. This computational efficiency might be crucial, most notably when the data is scarce, and uncertainty about the pandemic is abounding at the start of the pandemic. Our framework belongs to the observation-driven models class, specifically the GAS models proposed by \cite{creal2013}. GAS models involve many celebrated econometric models like the Generalized Autoregressive Heteroskedasticity (GARCH) model and various variants as a specific case, and thus, they proved to be useful in both fitting and prediction. \cite{koopman2016predicting} provide a comprehensive analysis of these models' predictive power compared to parameter-driven models in many settings, including models with count data.  

Observation-driven models for count data are considered, in many cases, independent of the analysis of the Covid-19 pandemic. \cite{Davis-etal-2003} provide a comprehensive analysis of observation-driven models with a particular focus on data with (conditional) Poisson distributions. \cite{Ferland-etal-2006} derive an integer-valued analog of the GARCH model (IN-GARCH) using Poisson distribution instead of Gaussian distribution. \cite{Fokianos-etal-2009} consider the Poisson autoregression of linear and nonlinear forms like the IN-GARCH model as a specific case. \cite{Chen-etal-2016} extend the Poisson autoregression to allow for smooth regime switches in parameters. Our framework naturally extends these approaches to the epidemiological model framework for each of the core compartments of the SIRD model using a multivariate structure.         

The remainder of the paper is organized as follows. Section~\ref{sec:Model specification} describes the canonical SIRD model and introduces the SIRD model with time-varying parameters. Section~\ref{sec:Econometric inference} discusses econometric issues, including identifying model parameters and how to account for sample selection. This section further elaborates on our estimation strategy and the resulting simulation scheme. Section~\ref{sec:Empirical results} presents estimation results using full sample data from the US. In Section~\ref{sec:Real-TimePerformance}, we evaluate our model framework's real-time performance in capturing the pandemic's current stance and forecasting compared to frequently used competitors. Section~\ref{sec:Extension} discusses potential extensions of the model. Finally, we conclude in Section~\ref{sec:Conclusion}. 

\section{Model specification}
\label{sec:Model specification}

\subsection{The canonical model of the pandemic, the SIRD model}
	
We start our analysis by discussing the epidemiological model denoted as the SIRD model of \cite{kermack1927}. Specifically, the SIRD model categorizes a population into four classes of individuals representing four distinct states of the pandemic; Susceptible ($S(t)$), Infected ($I(t)$), Recovered ($Rc(t)$) and Death ($D(t)$) in period $t$. The susceptible group does not yet have immunity to disease, and individuals in this group have the possibility of getting infected. On the other hand, the recovered group consists of individuals who are immune to the disease, and finally, $D(t)$ represents individuals who have succumbed to the disease. The Susceptible-Infected-Recovered-Death (SIRD) model builds on the principle that a fraction of the infected individuals in the population, $\frac{I(t)}{N}$, can transmit the disease to susceptible ones, $S(t)$, with a (structural) infection rate of $\beta$ by assuming a quadratic matching in the spirit of gravity law, see \cite{Acemoglu-etal-2020} for details on alternative matching structures. Therefore, the number of newly infected individuals in the current period is $\beta S(t) \frac{I(t)}{N}$. The newly infected individuals, that is, confirmed cases, $C(t)$, should be deducted from the susceptible individuals in the current period. Meanwhile, in each period, a fraction $\gamma$  of the infected people recover from the disease, which reduces the number of actively infected individuals. Similarly, a fraction $\nu$ of the infected people have succumbed to the disease, further reducing the number of actively infected individuals\footnote{We note the difference between the term \textit{death rate} and the terms  \textit{case fatality ratio} or \textit{mortality rate}. While the case fatality ratio refers to the ratio of the (cumulative) number of deaths to the (cumulative) number of the infected individuals, the \textit{mortality rate} measures the proportion of deaths due to a specific disease among the entire population for a given period. On the other hand, the death rate, $\nu_t$, measures the portion of the actively infected population who succumbed to Covid-19 for a given period.}. Hence, a fraction $\gamma + \nu$ of the infections are `resolved' in total. This structure leads to the following sets of equations: \vspace{-0.2cm}
\begin{equation}  \label{eq:SIR}
	\begin{array}{rcl}
\dot{S}(t)  & = & -\beta S(t) \frac{I(t)}{N} \\
\dot{Rc}(t) & = & \gamma I(t) \\
\dot{D}(t)  & = & \nu    I(t) \\
\dot{I}(t)  & = & \dot{S}(t) + \dot{Rc}(t) + \dot{D}(t)
%\dot{I}(t)  & = & \beta S(t) \frac{I(t)}{N} -  (\gamma + \nu)I(t),
	\end{array}	
\end{equation}
where $\dot{x}$ corresponds to $dx/dt$, and we assume that the population remains constant.\footnote{In fact, the number of deaths reduces the total population. We assume that the total number of deaths is negligible compared to the population for the tractability of the resulting SIRD model.} 

\subsection{Econometric analysis of the SIRD model with fixed parameters}
\label{subsec:FixedSIRDmodel}
The parameters of interest are the structural parameters $\beta$, $\gamma$, and $\nu$ that provide information on the transmission and resolution rates of the Covid-19 pandemic. A central metric that characterizes the course of the pandemic is the effective reproduction number, $eR(t)$. The effective reproduction number refers to the speed of the diffusion, which can be computed by the ratio of newly confirmed cases, denoted as $\dot{C}(t)$, to the resolved cases, that is, $\dot{C}(t) /(\dot{Rc}(t) + \dot{D}(t))$. Therefore, it serves as a threshold parameter of many epidemiological models for examining whether the disease will be extinct or spread further. Accordingly, using \eqref{eq:SIR} $eR(t)$ is identical to $\beta \frac{S(t)}{N} / (\gamma+\nu)$ and when $t=0$, it is identical to $\beta / (\gamma+\nu)$, in which case it is denoted as the basic reproduction rate. In this sense, a value of $eR(t)$ being less than unity indicates that the pandemic is contained, and if it exceeds unity, this implies that the spread of the pandemic continues. Our primary motivation for employing the model from the econometrics perspective is to conform to this canonical epidemiological model with the existing datasets and pinpoint the pandemic's stance timely. For that purpose, we first discretize \eqref{eq:SIR} as the typical Covid-19 dataset involves daily observations on the counts of individuals belonging to these states of health. Motivated by this, we specify a counting process for the states using the Poisson distribution conditional on past cases of active infections implying a nonhomogenous Poisson process for all the counts see, for example, \cite{Allen-2008, Yan-2008, Rizoiu-2018} for earlier examples, and \cite{li2020substantial} in the Covid-19 context for a similar approach. We specify the following for the stochastic evolution of the counts of these states;
	\begin{equation} \label{eq:FPSIR}
	\begin{array}{rcl}
	\Delta C_{t}|\Omega_{t-1} &\sim&  Poisson( \beta \frac{S_{t-1}}{N}I_{t-1}) \\
	\Delta Rc_{t}|\Omega_{t-1} &\sim&  Poisson( \gamma I_{t-1})    \\
	\Delta D_{t}|\Omega_{t-1} &\sim&  Poisson( \nu I_{t-1})    \\
    \Delta I_{t}   &=&  \Delta C_t - \Delta Rc_{t} - \Delta D_{t},
	\end{array}	
	\end{equation}	
where $\Omega_{t}$ stands for information set that is available up to time $t$. We assume that $\Delta C_t$, $\Delta Rc_t$, and $\Delta D_t$, representing the daily counts of the pandemic states, are independent conditional on $\Omega_{t-1}$. The final identity leads to an autoregressive process for the number of active infections, $I_t$. The resulting distribution for the number of active infections is a Skellam distribution (conditional on $\Omega_{t-1}$) with the mean $\pi_{t-1} I_{t-1}$, where $\pi_{t-1} = (1 +  \beta(1 -eR_{t-1}^{-1}))$ and the variance as $\beta(1+eR_{t-1})I_{t-1}$. Here, we use the identity in the last equation of \eqref{eq:FPSIR} together with the definition of $eR_{t}$. Therefore, the stationarity of the resulting process depends on whether $eR_{t-1}<1$ or $eR_{t-1}\ge 1$, i.e., whether the pandemic is taken under control or not. In addition, in case $\frac{S_{t-1}}{N} \approx 1$ or in other words, $eR_{t-1} \approx R_0$, the first and second unconditional moments are as follows, 
	\begin{equation} \label{eq:FPSIR-UnconMom}
	\begin{array}{rcl}
E[I_t] &=& \pi^t I_0  \\
Var(I_t) &=& \beta(1+R_0^{-1})\frac{\pi^{t-1}(1-\pi^{t})}{1-\pi} I_0,
	\end{array}	
	\end{equation}	
where we assume that the initial condition, $I_0$, is known. If the initial condition is considered a parameter to be estimated, then the variance is further amplified with a factor in the initial condition's variance. Accordingly, the unconditional moments of the pandemic states are linear functions of these unconditional moments of $I_t$. We refer to Section A of the supplementary material for details.

\subsection{Motivation for time variation in parameters}
\label{subsec:Motivation}
The canonical model's structural parameters represent the characteristics of the Covid-19 virus in terms of infectiousness and effectiveness. Therefore, unless there is a change in these characteristics, such as the emergence of a virus variant, the pandemic's underlying structural parameters remain unaffected. However, when confronting the epidemiological model with the real datasets of the pandemic, often, it is not feasible to observe/measure all the categories/compartments of the pandemic precisely. These measurement problems might arise from various reasons. For example, non-pharmaceutical interventions might be one of the underlying reasons, as identifying the number of perfectly isolated people who comply with stay-at-home orders is not easy. Even in a full lockdown scenario, some citizens might either break the rules or work in essential sectors that are always kept open. In addition, some behavioral shifts might be reflected in these parameters because, confronted with a high number of infections or alerted by the strict public measures, people would self-isolate even more with the fear of getting infected. These swings in attitudes are also reflected as time variations in parameters. Therefore, some authors refer to this distinction by rephrasing these as 'behavioral' parameters, see for example \cite{arias2021bayesian} or 'potential endogeneity of parameters', see for example \cite{avery2020economist}. Here we refer to a broader stance and refer to these parameters as 'implied' parameters in the sense that throughout the paper, the term 'parameter' refers to 'implied parameters'. Another motivation for the time-variation is the emergence of the new variant(s) of the virus with altered characteristics, such as the Delta and Omicron variants, for example, see \cite{karim2021omicron}. With the increasing number of infections by the new variant, the number of confirmed cases would correspond to a mixture of the prevailing variants. The weights of this mixture gradually evolve depending on the prevalence of more infectious variants. To elaborate further, we reconsider the first equation of the SIRD model in \eqref{eq:FPSIR}, i.e., the equation concerning the number of confirmed cases, this time considering the measurement error and the virus variants explicitly as
\begin{equation}
\label{eq:S-agentbased}
    \begin{array}{rcl}
    \Delta C_{t} & \propto & \frac{S^*_{t-1} }{N} (\beta_1 I_{1,t-1} + \beta_2 I_{2,t-1})
    \end{array}
\end{equation}
where $v=1,2$ denotes the $v^{th}$ variant of the virus and $I_{v,t-1}$ are the number of people infected with this variant. For ease of demonstration, we assume that in a given period $t$, only two variants can be spread in the population. $S^*_{t-1}$ indicates the number of susceptible people. However, we observe the number of susceptible people only with a measurement error that we denote with $\Xi_t$. Therefore, the observed number of susceptible people is $S_{t-1}=S^*_{t-1}+\Xi_t$. Moreover, the total number of active infections is the sum of the number of variant-specific infections over the type of infections, i.e., $I_t = I_{1,t} + I_{2,t}$. Considering the following decomposition, we can define the time-varying parameter as
\begin{equation}
\label{eq:TVP-beta-MeasErr}
    \begin{array}{rcl}
        \beta_t  \frac{S_{t-1}}{N}I_{t-1} &\approx& \frac{S^*_{t-1} }{N} (\beta_1 I_{1,t-1} + \beta_2 I_{2,t-1}) \\
          &=& \frac{S^*_{t-1} }{N} \left(  (\beta_2- \beta_1) I_{2,t-1} + \beta_1 I_{t-1} \right)   \\
          &=& \frac{S_{t-1} - \Xi_{t-1}}{N} \left(  (\beta_2- \beta_1) I_{2,t-1} + \beta_1 I_{t-1} \right) \\
           &=& \frac{S_{t-1}}{N}\beta_1 I_{t-1} + \frac{S_{t-1}}{N} (\beta_2- \beta_1) I_{2,t-1}  - \frac{\Xi_{t-1}}{N}  \beta_1 I_{t-1} - \frac{\Xi_{t-1}}{N}(\beta_2- \beta_1) I_{2,t-1}
      \end{array}
\end{equation}
The last three terms on the right-hand side represent various sources of factors that can lead to time variation in $\beta$. First, if in society only the first variant of the virus with the infection rate $\beta_1$ prevails and susceptible people are counted perfectly, then these terms drop from the expression and $\beta_t=\beta_1$. However, if a second variant of the virus emerges with the infection rate $\beta_2$ then we would observe a smooth change in the parameter with the increasing number of infections $I_{2,t-1}$ relative to $I_{1,t-1}$. If the dominance of the second variant does not materialize immediately, then the changes will be smooth until $\beta_t=\beta_2$, where the prevailing variant will be the second variant. On the other hand, if the number of susceptible could not be efficiently measured, then we would have a nonzero $\Xi_{t-1}$. This measurement error would magnify these changes further as, in this case, the third and fourth terms would contribute to the changes in the implied parameter. We provide a detailed analysis of these underlying causes of time variation related to measurement errors especially using the vaccination dataset and additionally in case of the probability of reinfection in Section B of the supplementary material. We refer to that section for a more detailed analysis. In this analytical demonstration, we only focus on the rate of infection for the discussion's compactness but modeling the underlying drivers of the rate of recovery and death could be conducted similarly. Essentially, for the remaining parameters, i.e., the rates of recovery and death, these arguments related to the difficulties in measurement and time-varying weight of the mixture of virus variants in the society might play an integral role in the time variation in parameters. In addition, advancements in the fight against Covid-19, including recovery of drugs and vaccination, could effectively mitigate the course of the disease.

\subsection{SIRD model with time-varying parameters - the TVP-SIRD model}
\label{subsec:TVPSIRDmodel}
In this section, we put forward the SIRD model with time-varying parameters. We use the framework of the Generalized Autoregressive Score model for modeling the time variation in parameters. This framework encompasses a wide range of celebrated models in econometrics, including the GARCH model and its variants. Briefly, the GAS model relies on the intuitive principle of modeling the time variation in key parameters in an autoregressive manner which evolves in the direction implied by the score function and thereby improving the (local) likelihood; see \cite{creal2013} for a detailed analysis of the GAS model. As in the case of the GARCH model, it effectively captures the time dependence in long lags in a parsimonious yet quite flexible structure. Consider the SIRD model with time-varying parameters as $\beta_t$, $\gamma_t$, and $\nu_t$. While the parameter for the rate of infection  $\beta_t$ is positive, the parameters $\gamma_t$ and $\nu_t$ are on the unit line. Therefore, we transform $\beta_t$ using logarithmic transformation and $\gamma_t$ and $\nu_t$ using logit transformation.\footnote{Essentially, we require an additional requirement that $\gamma_t+\nu_t \in [0,1]$. We go through a detailed specification for the convenient transformation of the parameters and consider three cases. In the first case, all three parameters are subject to only logarithmic transformation. In the second case, the parameters $\gamma_t, \nu_t \in [0,1]$ using logistic transformation, while in the third case, we impose an additional restriction of $\gamma_t+\nu_t \in [0,1]$. Results show that the final restriction  $\gamma_t+\nu_t \in [0,1]$ is not binding. However, it imposes important challenges on the feasibility of the estimation, especially when we enhance the model to allow for seasonality. Therefore, we proceed with the second case throughout the paper. The specification search and the findings are displayed in Section C of the supplementary material.} Let the parameter with a $\tilde{(.)}$ denote the corresponding transformations as $\tilde{\beta}_t = \ln(\beta_t)$, $\tilde{\gamma}_t = \text{logit}(\gamma_t)$ and $\tilde{\nu}_t = \text{logit}(\nu_t)$ where logit refers to the inverse of the logistic transformation as $\text{logit}(x) = \log(\frac{x}{1-x})$. The resulting Time-Varying Parameters - SIRD (TVP-SIRD) model is as follows
\begin{equation}
\label{eq:TVP-SIRD}
    \begin{array}{rcl}
    \Delta C_{t}|\Omega_{t-1} &\sim&  Poisson( \beta_t \frac{S_{t-1}}{N}I_{t-1}) \\[-0.2em]
	\Delta Rc_{t}|\Omega_{t-1} &\sim&  Poisson( \gamma_t I_{t-1})    \\[-0.2em]
	\Delta D_{t}|\Omega_{t-1} &\sim&  Poisson( \nu_t I_{t-1})    \\[0.4em]
    \Delta I_{t}   &=&  \Delta C_t - \Delta Rc_{t} - \Delta D_{t}.\\
    \end{array}
\end{equation}
We decompose the transformed parameters further into a smooth level component and a high-frequency seasonal component. Because the typical daily Covid-19 dataset exhibits an immense daily seasonal pattern potentially due to frictions in reporting, we also put a particular emphasis on the seasonal component. Specifically, consider the decomposition as
\begin{equation}
\label{eq:TVP-SIRD-decompose}
    \begin{array}{rcl}
    \theta_t & = &  \theta_{l,t} + \theta_{s,t} 
    \end{array}
\end{equation}
where for parameter $\theta_t = \tilde{\beta}_t, \tilde{\gamma}_t$ and $\tilde{\nu}_t$, respectively, $\theta_{l,t}$ and  $\theta_{s,t}$ are the level and seasonal components, respectively.\footnote{Such additive structure in the transformed parameters leads to a multiplicative seasonal structure in exponential form as a function of original parameters; see, for example, \cite{hansen2021dynamic} for a similar approach.} For the level parameter, the evolution of the parameters is specified as
\begin{equation}
\label{eq:TVP-SIRD-level}
    \begin{array}{rcl}
% \theta_{l,t}   &=&  \theta_{l,t-1}   + \alpha_{\theta} s_{\theta, t-1} 
\theta_{l,t}   &=& \omega_{\theta} + \beta_{\theta} \theta_{l,t-1}   + \alpha_{\theta} s_{\theta, t-1} 
   \end{array}
\end{equation}
where $s_{\theta,t}$ for $\theta_t = \tilde{\beta}, \tilde{\gamma}_t$ and $\tilde{\nu}_t$ are the (scaled) score functions of the joint likelihood which is identical to that for the level parameter due to the additive structure. 
Since the SIRD model's likelihood function is constituted by the (conditionally) independent Poisson processes, each score function is derived using the corresponding compartment. Specifically, let   $\nabla_{\tilde{\beta},t}=\frac{\partial L(\Delta C_{t}; \tilde{\beta}_t)}{\partial \tilde{\beta}_t}$, $\nabla_{ \tilde{\gamma},t}=\frac{\partial L(\Delta Rc_{t}; \tilde{\gamma_t})}{\partial  \tilde{\gamma_t}}$ and $\nabla_{ \tilde{\nu},t}=\frac{\partial L(\Delta D_{t}; \tilde{\nu_t})}{\partial  \tilde{\nu_t}}$ denote the score functions for period $t$ observation. We specify $s_{\theta,t}$ such that the score functions are scaled by their variance as $s_{\theta,t}=\frac{\nabla_{\theta,t}}{\text{Var}(\nabla_{\theta,t})}$  for $\theta_t = \tilde{\beta}, \tilde{\gamma}_t$ and $\tilde{\nu}_t$.\footnote{Alternative approaches for scaling the score function include the standard deviation rather than the variance and the score function without scaling. Our findings suggest that using the variance as the scaling function leads to smoother and more robust evolution of parameters over time.}  In the specific case of the SIRD model, this modeling strategy leads to the following specification for the (scaled score functions) in terms of the corresponding link function
	\begin{equation}
	\label{eq:ScoreFunctions}
        \begin{array}{rcl}
	    s_{\tilde{\beta},t} &=& \frac{\Delta C_{t-1} - \lambda_{1,t-1}}{\lambda_{1,t-1}}  \\
    s_{\tilde{\gamma},t} &=& \frac{\Delta R_{t-1} - \lambda_{2,t-1}}{\lambda_{2,t-1}}\frac{1}{(1-\gamma_t)} \\
    s_{\tilde{\nu},t} &=& \frac{\Delta D_{t-1} - \lambda_{3,t-1}}{\lambda_{3,t-1}} \frac{1}{(1-\nu_t)}	   
    	\end{array}
     \end{equation} 
where $\lambda_{1,t} = \beta_t \frac{I_{t-1}S_{t-1}}{N}$, $\lambda_{2,t} = \gamma_t I_{t-1}$ and $\lambda_{3,t} = \nu_t I_{t-1}$. The resulting specification implies an intuitive updating rule because the parameters (in the logarithmic form) are updated using a combination of the previous parameter value and a function of the previous percentage deviation from the mean. We refer to Section D of the supplementary material for the details on the derivation of \eqref{eq:ScoreFunctions}. One drawback of the specification in \eqref{eq:TVP-SIRD-level} is that when $\beta_{\theta}$ is close to unity, identification of $\omega_{\theta}$ is cumbersome, see for example \cite{kastner2014ancillarity}. This is also our experience when estimating the model using real datasets. Therefore, we restrict the $\beta_{\theta}$ parameter to be unity. In this case, we remove the intercept parameter $\omega_{\theta}$, but the level of the time-varying parameters is estimated as the initial condition $\theta_{l,0}$  along with other model parameters.\footnote{We also compute the Bayes factor of this model relative to the unrestricted model for formal model comparison. We find that the Bayes factor is to be 0.99, indicating that the restriction is also supported by the data, as expected.} 

For the seasonal component, we specify a structure using frequency domain for capturing the daily seasonality in a given week adequately departing from \cite{hansen2021dynamic}\footnote{We also consider models for seasonality that exploits the time domain using daily components for modeling seasonality where the corresponding coefficients are time-varying. Our findings suggest that models in the frequency domain facilitate the estimation and performs better than their time domain counterpart. We display these results in Section E of the supplementary material.}. 
Consider the model
\begin{equation}
\label{eq:TVP-SIRD-Seas}
\begin{array}{rcl}
\theta_{s,t} &=&  \sum_{j=1}^{s/2}  \theta_{js,t}  
\end{array}
\end{equation}
with 
\begin{equation}
\label{eq:TVP-SIRD-Seas-Parts}
\begin{array}{rcl}
\theta_{js,t}  &=& \cos(\Lambda_j)  \theta_{js,t-1} + \sin(\Lambda_j) \theta^*_{js,t-1} + \psi_j s_{\theta, t-1}  \\
\theta^*_{js,t}  &=&  -\sin(\Lambda_j) \theta_{js,t-1} + \cos(\Lambda_j)\theta^*_{js,t-1} + \psi_J^* s_{\theta, t-1}  \end{array}
\end{equation}
where $\Lambda_j=\frac{2\pi j}{s}$ for $j=1,2,3$ and $\theta_t = \tilde{\beta}, \tilde{\gamma}_t$ and $\tilde{\nu}_t$. 
The structure in \eqref{eq:TVP-SIRD-Seas} and \eqref{eq:TVP-SIRD-Seas-Parts} provides a quite flexible yet parsimonious model structure for capturing seasonal behavior. Essentially, it can be shown that in case the score functions are zero, it reduces to
\begin{equation}
\label{eq:Modulate}
\begin{array}{rcl}
\theta_{js,t}  &=&  -\sin(\Lambda_j t) \theta_{js,t-1} + \cos(\Lambda_j t)\theta^*_{js,t-1}. 
\end{array}
\end{equation}
This structure implies that the cycle is captured by the three periodic series with frequencies $\Lambda_1=\frac{2\pi}{7}, \Lambda_2=\frac{4\pi}{7}$ and $\Lambda_3=\frac{6\pi}{7}$ each with period 7, 3.5 and 2.3 days. While the first series has the fundamental frequency, the remaining parts could be obtained by integrating the fundamental frequency; see \cite{proietti2022seasonality} for further details. The sine and cosine terms together function as two orthogonal bases generalizing the model in \cite{hansen2021dynamic}. As in the level case, the score function is identical to the general score function owing to the linear decomposition of the parameters into the level and seasonal components. This structure facilitates the estimation substantially and enables us to capture the potential link between level and seasonality. 

The specification in \eqref{eq:TVP-SIRD}-\eqref{eq:TVP-SIRD-Seas-Parts} leads to quite rich dynamics both in terms of mean and the variance of the resulting process. These rich dynamics enable us to accurately capture the pandemic's evolution reflected in the timely and prompt response of the parameters to the pandemic states' data changes. To elaborate further, we also consider the implied moments of the resulting process. We display these findings in Section F of the supplementary material.

\section{Econometric inference}
\label{sec:Econometric inference}
Two major issues plague the inference of the SIRD model parameters. First, from the epidemiology point of view, the SIRD model could be extended in various directions by incorporating other phases, or in other words, compartments of the disease. As this implies additional parameters to be estimated, identifying these parameters is challenging. Second, detection of the infected individuals might be burdensome as some of them do not show symptoms, yet they are infectious. In this section, we first discuss the challenges of the econometric inference, and second, we introduce the details of the simulation-based Bayesian estimation strategy for the econometric inference of the TVP-SIRD model.

\subsection{Identification of model parameters}
\label{subsec:Identification}
The pandemic's course in the evolution of active cases depends on the structural parameters, $\beta, \gamma$ and $\nu$ that are used to construct $\pi$ in \eqref{eq:FPSIR-UnconMom}. The initial condition  $I_0$ is also required because the process might be nonstationary if the pandemic is not contained. We estimate the models starting from the period when the number of cumulative confirmed cases exceeds 1000, and we use this first observation in our sample as the initial condition.\footnote{Different starting points (such as the periods when the number of cumulative confirmed cases exceeds 10000) yield very similar results. Results are available upon request by the authors.} The structural parameters represent the compartments of the SIRD model where the compartments refer to the specific phases of the disease as 'susceptible', 'infected', 'recovered', or 'death'. Still, it is possible to extend the model with additional compartments. For example, it is known that the virus has an incubation period in which the susceptible person is 'exposed' to the virus but not yet affected by it. Nevertheless, she can transmit the virus to other people. Departing from this point \cite{KOROLEV2020},  for example, discusses identification problems of the structural parameters regarding the SIRD model with an additional compartment of 'exposed' case. However, these compartments require additional parameters to be estimated; see \cite{lourenco2020fundamental} for details. If the specific compartments' data, such as the number of infected cases where the virus is in the incubation period, is available, these structural parameters are well identified. Therefore, while these additional compartments provide further refinements to the SIRD model, these refinements plague the identification of the structural parameters if additional data is missing; see, for example, \cite{Atkeson-2020} who discusses the identification of the structural parameters regarding the SIRD model. He demonstrates how different parameter setups might result in very similar initial phases of the pandemic but result in divergent patterns in the long-run absence of the data on the model's compartments. 

\subsection{Accounting for sample selection}
	\label{subsec:Asymptomatics}
A fundamental underlying assumption of the model specification in previous sections is that the variables of the infected, recovered, and succumbed individuals represent the aggregate numbers. However, one of the stylized facts related to the Covid-19 pandemic is the presence of infected individuals who do not have any symptoms, denoted as asymptomatic. These hard-to-detect cases complicate the analysis as it leads to a selection bias in econometric inference, among other factors, see \cite{manski2020estimating}. These unreported infection cases prohibit the tests from being randomly assigned, plaguing econometric inference. This section provides a model extension based on some assumptions on the model structure to capture asymptomatic infected individuals. We use two sources of additional datasets to extract the total number of active cases, including the reported and unreported ones. The first additional data we use is the number of excess deaths. These include the estimates of additional deaths directly or indirectly attributed to Covid-19 in excess of the published number of deaths. This projection provides {\bf weekly} estimates of excess deaths, and these weekly counts of deaths are compared with historical trends. Accordingly, it provides an essential source of identification for the unreported cases, as potentially a significant part of the excess deaths might be attributed to this sample selection\footnote{Please see, \url{https://www.cdc.gov/nchs/nvss/vsrr/covid19/excess_deaths.htm} and \url{https://ourworldindata.org/excess-mortality-covid} for additional discussion and methodology on the computation of excess death}. The second variable we exploit is the number of positives in the tested individuals. To motivate the idea, let $P_t$ denote an indicator function that takes the value one if an individual is infected in period $t$ and 0 otherwise. Further, let $T_t$ denote another indicator function, which takes the value one if an individual is tested for the infection in period $t$ and 0 otherwise. Using the Bayes rule, we can show that, 
\begin{equation}
    P(P_t=1) = \frac{P(P_t=1|T_t=1)P(T_t=1)}{P(T_t=1|P_t=1)},
\end{equation}
see also \cite{stock2020}. In case the assignment for testing of an individual is carried out randomly, then $P(T_t=1)=P(T_t=1|P_t=1)$ and there is no identification problem due to sample selection. Neither, $P(P_t=1)$ nor $P(T_t=1|P_t=1)$ are observed. Nevertheless, $P(T_t=1)$ could be computed as the fraction of tested individuals in the population in period $t$. Furthermore, $P(P_t=1|T_t=1)$ could be considered as the daily positive test rate. Equipped with these, identification of $P(T_t=1|P_t=1)$ boils down to the identification of $P(P_t=1)$, the true prevalence of the infection, including asymptomatic cases. Departing from \cite{Grewelle-2020}, we make use of a parametric identification strategy for approximation of $P(T_t=1|P_t=1)$,
\begin{equation}
\label{eq:Approx}
    P(T_t=1|P_t=1) = \exp(-k\rho_t)
\end{equation}
where $\rho_t$ is the fraction of positives in the tested individuals in period $t$, and $k$ is a positive constant. Briefly, the underlying idea stems from the fact that detecting infections, including asymptomatic individuals, would improve with the increasing number of testing. In that sense, the fraction of tested individuals in the population should be related to the ratio of reported infections to the total number of infections. With the increasing number of testing on the population, this fraction approaches one. On the other hand, if testing is concentrated only on symptomatic individuals, this fraction approaches a lower bound, captured by the parameter $\exp(-k)$, where the functional form admits exponential decay.

Unlike the daily dataset we employ for estimating the TVP-SIRD model, the excess death data is at the weekly frequency. Therefore, we extend our model framework to allow for a mixed-frequency dataset. Let $I^*_{t}$ be the number of infected individuals involving asymptomatic and symptomatic cases, and let $S^*_t$ and $Rc^*_t$ denote the total number of susceptible and recovered individuals, respectively. Let $\delta_t = 1-\exp(-k\rho_t)$ denote the fraction of the unreported infection cases among all infection cases. Using \eqref{eq:Approx}, these could be computed as
\begin{equation}
\begin{array}{rcl}
    \Delta C_t^* &=&  \frac{\Delta C_t}{1-\delta_t} \\
   \Delta Rc_t^*  &=& \frac{\Delta Rc_t}{1-\delta_t} \mbox{  for $t=1,2, \dots, T$}. 
\end{array}
 \end{equation}
Further denote the total number of weekly deaths as $\Delta \bar{D}^w_{t}$ which is computed as, 
\begin{equation}
\begin{array}{rcl}
   \Delta \bar{D}^w_{t}  &=&  \Delta D^w_{t} + \Delta ED_{t} \mbox{  for  $t=7k$ and $k=1,2,\dots$ }, 
\end{array}
 \end{equation}
 where $D^w_{t}$ stands for the reported deaths at the weekly frequency and $\Delta ED_{t}$ for the Excess Death numbers estimated in period $t$.  
Finally, the mixed frequency TVP-SIRD (MF-TVP-SIRD) model in terms of the total numbers can be written as
\begin{equation}
\label{eq:MF-TVP-SIRD}
    \begin{array}{rcl}
     \Delta C^*_{t}|\Omega_{t-1} &\sim&  Poisson( \beta_t \frac{S^*_{t-1}}{N} I^*_{t-1}) \\[-0.2em]
     \Delta Rc^*_{t}|\Omega_{t-1} &\sim&  Poisson( \gamma_t  I^*_{t-1} )  \mbox{  for $t=1,2, \dots, T$}   \\[-0.2em]
   	   \Delta \bar{D}^w_{t}|\Omega_{t-1} &\sim&  Poisson( \sum\limits_{s=t-6}^{t} \nu_s I^*_{s-1})   \mbox{  for  $t=7k$ and $k=1,2,\dots$ }  \\[0.4em]
    \Delta C^*_t = -\Delta S^*_t    &=&  \Delta I^*_{t} + \Delta Rc^*_{t} + \Delta \bar{D}^{d}_{t} \\
    \end{array}
\end{equation}
Here $\Delta \bar{D}^d_{t}$ is computed as $\nu_t I^*_{t-1}$\footnote{Notice that the sum of the independent Poisson processes is a Poisson process, which enables us to switch between daily and weekly frequency when necessary.}.
The evolution of the model parameters decomposed as in \eqref{eq:TVP-SIRD-decompose} follow the recursions in \eqref{eq:TVP-SIRD-level} and \eqref{eq:TVP-SIRD-Seas} using the (scaled) score functions displayed in \eqref{eq:ScoreFunctions} for the parameters $\beta_t$ and $\gamma_t$. For the score function of the $\nu_t$, we have the following expression due to the change in the frequency of the Poisson process
\begin{equation}
\label{eq:WeeklyScore}
    \begin{array}{rcl}
    s_{\tilde{\nu},t} &=& \frac{\Delta \bar{D}_{t} - \bar{\lambda}_{3,t-1}}{\bar{\lambda}_{3,t-1}} \frac{1}{(1-\nu_t)}
    \end{array}
\end{equation}
with $\bar{\lambda}_{3,t-1} = \nu_t \sum\limits_{s=t-6}^{t} I^*_{s-1}$  for  $t=7k$ and $k=1,2,\dots$, that is for the periods where the weekly data is released, i.e., observed. The score function takes the value 0 when the weekly excess death variable is not observed, which completes the specification of the MF-TVP-SIRD model. 

\subsection{Estimation strategy and the simulation algorithm} 
\subsubsection{Bayesian inference}
We use simulation-based Bayesian estimation techniques for inference on model parameters. Bayesian inference involves updating the prior distributions of model parameters with the data likelihood to form the parameters' posterior distributions. Considering the SIRD model, Bayesian inference is especially appealing since the inference is conditional on the data at hand and does not require asymptotic analysis. Therefore, we can compute the credible intervals when the underlying process of the number of infected cases, $I_t$, is nonstationary. This property is especially reassuring in our case since, obviously, nonstationarity, or in other words, effective reproduction rate being greater than one, $eR_{t}>1$, is an inevitable feature of the pandemic, at least locally. 

Here we demonstrate the likelihood function and prior specifications for the TVP-SIRD model because the SIRD model with fixed parameters boils down to a special case of this model. The likelihood function is based on the specification in \eqref{eq:TVP-SIRD} where we specify conditionally independent Poisson distributions for each of the components of the SIRD model. Let $y_t=(\Delta C_t, \Delta Rc_t, \Delta D_t)'$ be the vector of observations. Notice that $I_t =I_{t-1} + \Delta C_t -\Delta Rc_t - \Delta D_t$, and thus the number of active infections can be computed using the information set $\Omega_t = (y_t', I_{t-1})'$. Accordingly, we have the following likelihood function
\begin{equation}             
\label{eq:LikelihoodFunction}
       \begin{array}{rcl}
    f(y_t|\Omega_{t-1}) = \frac{\lambda_{1,t}^{\Delta C_t} \exp(-\lambda_{1,t})}{\Gamma(\Delta C_t + 1)} ~ \frac{\lambda_{2,t}^{\Delta Rc_t} \exp(-\lambda_{2,t})}{\Gamma(\Delta Rc_t + 1)} ~\frac{\lambda_{3,t}^{\Delta D_t} \exp(-\lambda_{3,t})}{\Gamma(\Delta D_t + 1)},
 	\end{array}
 \end{equation}
 where $\lambda_{1,t} = \beta_t \frac{S_{t-1}I_{t-1}}{N}$, $\lambda_{2,t} = \gamma_t I_{t-1}$  and $\lambda_{3,t} = \nu_t I_{t-1}$ and $\Gamma(.)$ is the Gamma function. Note that the time-varying parameters decomposed as in \eqref{eq:TVP-SIRD-decompose}  follow the recursions in \eqref{eq:TVP-SIRD-level} and \eqref{eq:TVP-SIRD-Seas} using the (scaled) score functions displayed in \eqref{eq:ScoreFunctions}. We want to obtain posterior results driven by the data rather than the prior distributions. Therefore, we impose rather diffuse prior specifications for the model parameters. This strategy implies that for the representative model parameters $\phi$, we specify the following improper prior specifications
\begin{equation}
\label{app:eq:Priors}
     \begin{array}{rcl}
    f(\phi) \propto 1
 	\end{array}
 \end{equation}
 for $\phi \in \Phi=(\Theta_{l,0}', \alpha', \psi^{'}_{\tilde{\beta}}, \psi^{*'}_{\tilde{\beta}}, \psi^{'}_{\tilde{\gamma}}, \psi^{*'}_{\tilde{\gamma}}, \psi^{'}_{\tilde{\nu}}, \psi^{*'}_{\tilde{\nu}})'$ where $\Theta_{l,0}=(\beta_{l,0}, \gamma_{l,0}, \nu_{l,0})'$, $\alpha=(\alpha_{\beta}, \alpha_{\gamma}, \alpha_{\nu})'$,   $\psi_{\theta}=(\psi_{1,\theta}, \psi_{2,\theta}, \psi_{3,\theta})'$ and $\psi^{*}_{\theta}=(\psi^{*}_{1,\theta}, \psi^{*}_{2,\theta}, \psi^{*}_{3,\theta})'$ for  $\theta_t = \tilde{\beta}, \tilde{\gamma}_t$ and $\tilde{\nu}_t$. For the MF-TVP-SIRD model, we specify the prior distribution for the additional $k$ parameter noninformative in the positive domain. 
 
 \subsubsection{Simulation scheme}
 For the SIRD model with fixed parameters, the likelihood with Poisson distributions as in \eqref{eq:LikelihoodFunction} with fixed parameters and noninformative or conjugate priors in the form of Gamma distribution lead to a Gamma distribution for the posterior distributions of the model parameters. Therefore, these can be sampled using the plain Gibbs sampler. For the TVP-SIRD model, the fact that we have time-varying parameters with deterministic recursions leads to nonstandard posterior distributions. Therefore, we cannot use standard distributions we can easily simulate for the inference, as is the case for the Gibbs sampler. Instead, we resort to the (adaptive) random walk Metropolis-Hastings (MH) algorithm within the Gibbs sampler; see \cite{robert2013monte} for details. The algorithm is as follows
\begin{enumerate}
    \item Sample $\alpha$ from $f(\alpha|S^T, I^T, \Phi_{-\alpha})$ using MH step
    \item Sample $\Theta_{l,0}$ from $f(\Theta_{l,0}|Rc^T, I^T, \Phi_{-\Theta_{l,0}})$ using MH step
    \item Sample $\psi_{\theta}$ from $f(\psi_{\theta}|Y^T, \Phi_{-\psi_{\theta}})$ using MH step
    \item Sample $\psi^{*}_{\theta}$ from $f(\psi^{*}_{\theta}|Y^T, \Phi_{-\psi^{*}_{\theta}})$ using MH step
\end{enumerate}
Here $Y^T= (Y_1,\dots,Y_t,\dots,Y_T)'$ for $Y_t=S_t,I_t,Rc_t,D_t$ indicating the full sample of the count data regarding the states of the pandemic, respectively. $\Phi_{-X}$ indicates the vector of parameters $\Phi$ excluding the parameters $X$. For the MH steps, the candidate generating density is constructed using the random walk specification as
\begin{equation}
\label{eq:RWMH}
\begin{array}{rcl}
\phi_m& = &\phi_{m-1} + \Sigma_\phi^{1/2} \varepsilon_m
\end{array}    
\end{equation}
where $\phi_m$ is the parameter draw depending on the step at the iteration $m$, and $\varepsilon_m$ follows a standard (multivariate) $t-$distribution with degrees of freedom 15. For the starting values of the parameters to initialize the sampler, $\Phi_0$, and for the covariance matrix $\Sigma_{\Phi_0}$, we use the maximum likelihood estimate of the model parameters. Therefore, we use the mode and the inverse Hessian of the likelihood function at the mode in \eqref{eq:RWMH}. To improve the sampler's performance, we follow the adaptive scheme described in \cite{haario2001adaptive}. This scheme involves replacing $\Sigma_{\Phi_0}$ with $\chi S_M + \epsilon I$, once we obtain a sufficient number of draws to replace the inverse Hessian of the likelihood function with the simulated curvature of the posterior distribution. Here $S_m$ corresponds to the empirical covariance matrix computed using the draws up to step $M$, $I$ indicates the identity matrix, and $\epsilon$ is a small number. $\epsilon I$ ensures a nonsingular empirical covariance matrix. In addition, we use $\chi$ for optimizing the sampler's performance for the candidate generating density to be efficient enough to cover the tails of the posterior distribution. Let $\phi^{cand} \sim q(\phi^{cand}|\phi_{m-1})$ be a draw from the candidate generating density in iteration $m$ of the sampler, the candidate is accepted with probability
\begin{equation}
\label{eq:AccProb}
\begin{array}{rcl}
\pi &=& \min \left\{ 1, \frac{q(\phi_{m-1}|\phi^{cand}) p(\phi^{cand}|Y^T)}{q(\phi^{cand}|\phi_{m-1})p(\phi_{m-1}|Y^T) }   \right\}. 
\end{array}    
\end{equation}
Here $p(.|Y^T)$ refers to the posterior distribution. Note that, due to the symmetry of the random walk specification, $q(\phi_{m-1}|\phi^{cand})$ and $q(\phi^{cand}|\phi_{m-1})$ are equivalent. Hence they are of no use in \eqref{eq:AccProb}.

\section{Empirical results}
\label{sec:Empirical results}
\subsection{Dataset}
We use the data for the US starting from the early days of the pandemic until the end of March 2022, which captures all major waves of infections related to the Covid-19 pandemic, including the recent Omicron wave. 
%For the full sample analysis, we use the dataset that is available as of the beginning of April. 
The data is originally published by the Center for Disease Control and Prevention (CDS) and can be tracked on \url{https://covid.cdc.gov/covid-data-tracker/\#datatracker-home}. The data on the number of recovered cases ceased to be reported following December 2020. We, therefore, treat this data as missing for the periods when the data is not reported. In these cases, the score function is set to zero, similar to the estimation of the MF-TVP-SIRD model. For the out-of-sample analysis that involves a real-time recursive prediction exercise, we use the US data vintages that are available as of the day of the prediction, which is available on the Covid-19 Data-Hub. The data-hub includes the daily vintages of Covid-19 pandemic related datasets; see \url{https://covid19datahub.io/articles/data.html} and \cite{guidotti2020} and \cite{guidotti2022} for implementation details and the latest version of the dataset. 

\subsubsection{A brief account of the pandemic's course in the US}
\begin{figure}[H]
    \centering
    \caption{The evolution of the daily number of confirmed and death cases in the US}
\begin{tabular}{cc}
 Confirmed &  Death\\[-0.2em]
  \includegraphics[trim = 0mm 6mm 0mm 0mm, clip,width=0.51\textwidth]{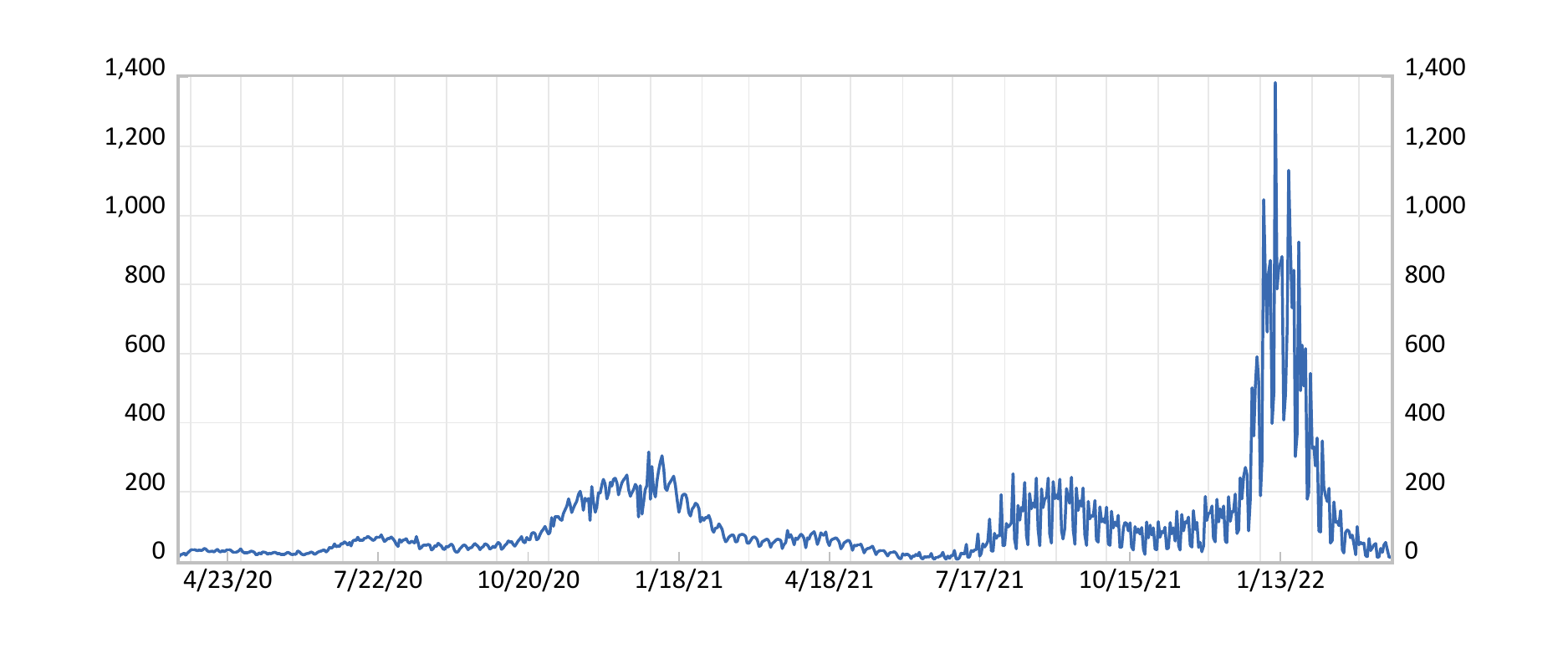}  
  &  \includegraphics[trim = 0mm 6mm 0mm 0mm , clip, width=0.51\textwidth]{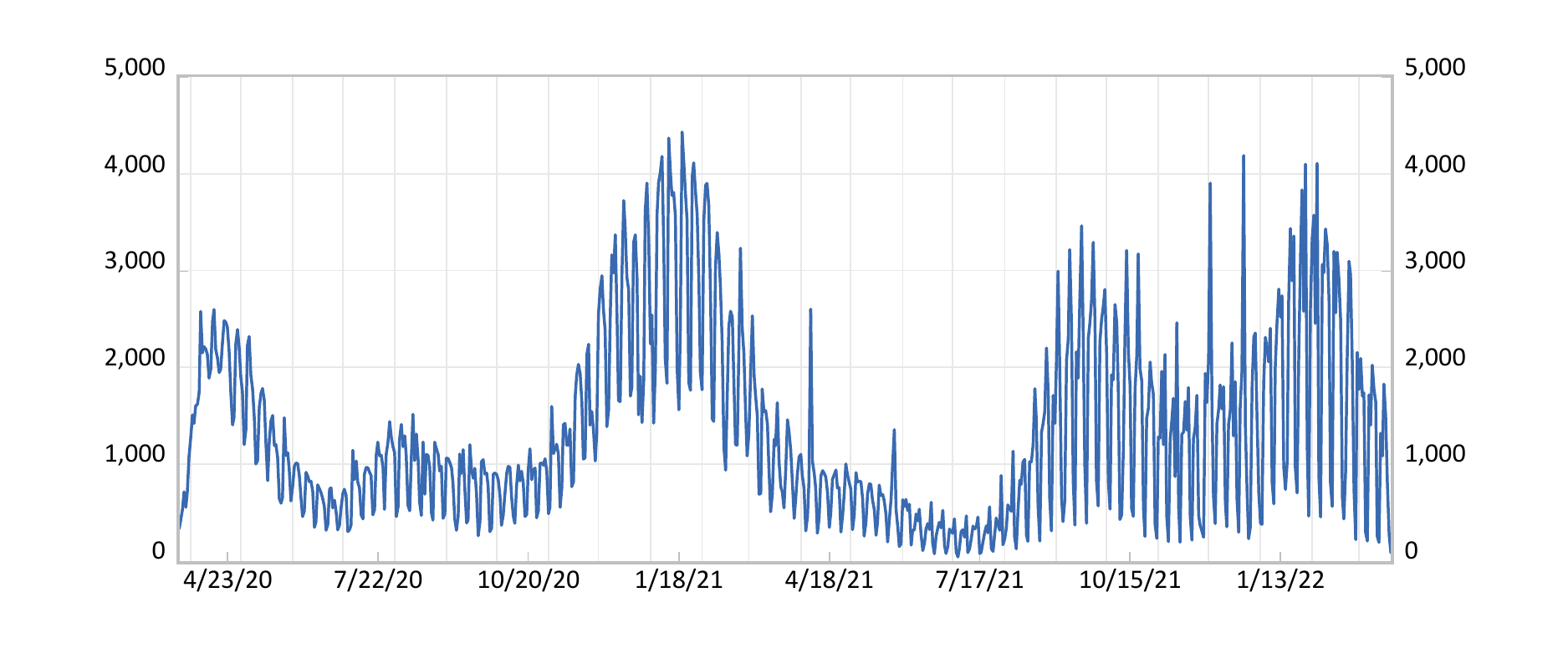} 
   \end{tabular}
    \label{fig:ConfirmedDeathCases}
\end{figure}

We display the evolution of the daily confirmed cases and deaths throughout the sample in Figure~\ref{fig:ConfirmedDeathCases}.
\begin{center}
[Insert Figure~\ref{fig:ConfirmedDeathCases} about here]
\end{center} 
The US exhibits extensive heterogeneity throughout the sample regarding their experience related to the pandemic, as shown in Figure~\ref{fig:ConfirmedDeathCases}. At the onset of the pandemic, the US opted for imposing mixed strategies involving full and partial lockdowns and voluntary quarantine in different regions. As of March 2020, the US reported the highest number of daily confirmed cases worldwide and therefore was the center of the pandemic when the country was struggling with the first wave. In late 2020, the Alpha variant was the virus's dominant strain, which can be detected as the second wave in Figure~\ref{fig:ConfirmedDeathCases}. Nevertheless, with the start of 2021, the US began an intensive inoculation campaign to fight the pandemic. In 2021 two major pandemic waves were experienced with the emergence of the Delta and Omicron variants, among other strains. While the Omicron variant notably led to record values for the number of daily confirmed cases, the daily number of deaths was still comparable to that of the Alpha variant owing to the success of vaccination efforts. Hence, this relatively rich and heterogeneous dataset involving all sorts of pandemic experiences enables us to examine the econometric model's success in tracking the parameter changes in response to policy implementations and changes in the virus's characteristics. 

\subsection{Full sample results}
\begin{figure}[H]
    \centering
    \caption{The evolution of the effective reproduction rate, $eR_t$, estimated using the FP-SIRD model}
\begin{tabular}{c}
   \includegraphics[trim = 3cm 10cm 3cm 10cm, clip, width=0.7\textwidth]{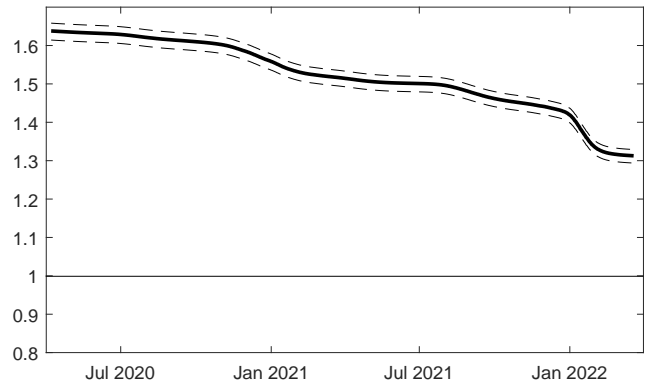}      \end{tabular}
    \label{fig:eRt_FPSIRD}
\end{figure}

This section discusses the full sample estimates of the main parameters for the models with fixed and time-varying parameters, as described in \eqref{eq:FPSIR} and \eqref{eq:TVP-SIRD}, respectively. We further evaluate the model's parameter estimates with time-varying parameters when asymptomatic cases are explicitly considered, as shown in \eqref{eq:MF-TVP-SIRD}.  
We start our analysis with the full sample estimates of the model parameters for the SIRD model with fixed parameters described in \eqref{eq:FPSIR}. These are displayed in Panel A of Table~\ref{tab:FP-SIR-Parameters-TVP_alpha}. 
\begin{center}
[Insert Table~\ref{tab:FP-SIR-Parameters-TVP_alpha} about here]
\end{center} 
The model with fixed parameters reflects the stance of the pandemic over the last two years, on average, as the parameters are kept fixed. The basic reproduction rate, $R_0$, as the main summary statistics on the course of the pandemic when the whole population is considered susceptible, is displayed in the last column of Table~\ref{tab:FP-SIR-Parameters-TVP_alpha}. The median estimate of the $R_0$ for the US shows that over the sample period of more than two years since the start of the pandemic in March 2020, the estimate is 1.64 with little uncertainty. The fact that the $R_0$ well exceeds 1 reflects that the pandemic is not contained yet on average with the repeated waves. Notice that this value might be inaccurate because the number of susceptible people has reduced to some extent with the progress of the pandemic. We, therefore, display the evolution of the effective reproduction rate, $eR_t$ in Figure~\ref{fig:eRt_FPSIRD}. 
\begin{center}
[Insert Figure~\ref{fig:eRt_FPSIRD} about here]
\end{center}
Figure~\ref{fig:eRt_FPSIRD} indicates that the effective reproduction rate declines over time with the reduced number of susceptible people. As the main reason for the reduction in the number of susceptible people is the infected cases, the reduction in $eR_t$ closely follows the waves of the pandemic. The final value as of the end of March 2022 is still 1.3, indicating that a large part of the population is not infected yet. 

\begin{figure}[H]
    \centering
    \caption{The fitted values of the daily number of confirmed and death cases using the TVP-SIRD model}
\begin{tabular}{cc}
 Confirmed &  Death\\[-0.2em]
  \includegraphics[trim = 0mm 6mm 0mm 0mm, clip,width=0.51\textwidth]{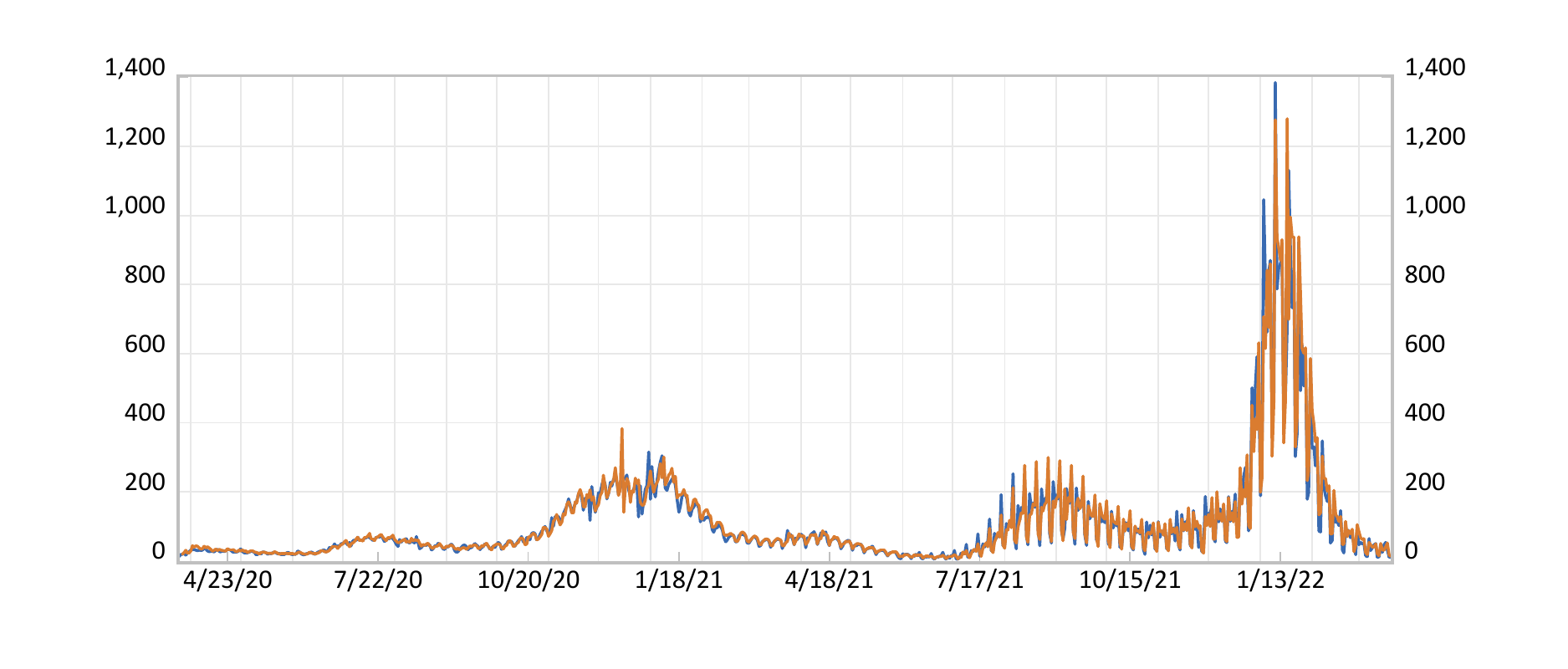}  
  &  \includegraphics[trim = 0mm 6mm 0mm 0mm , clip, width=0.51\textwidth]{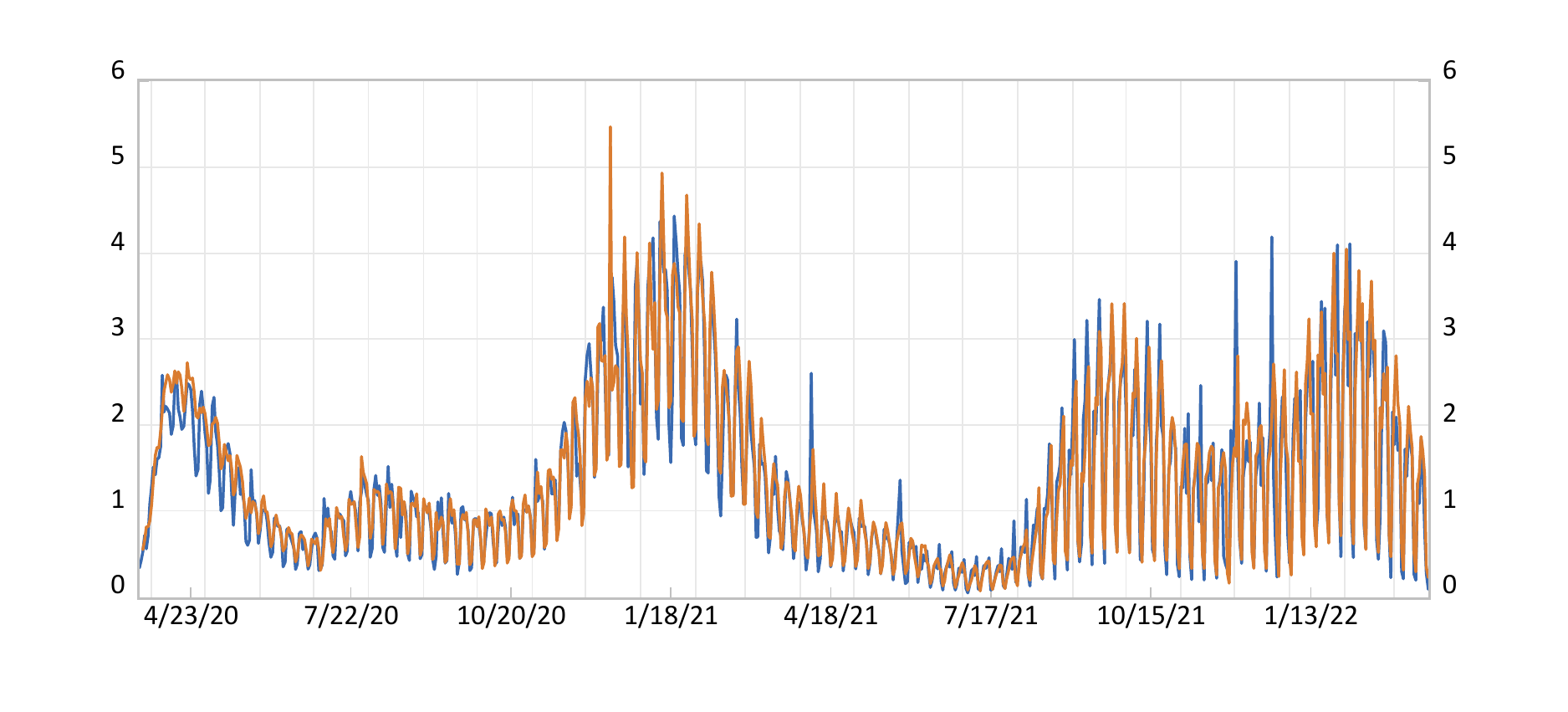} 
   \end{tabular}
    \label{fig:FittedConfirmedDeathCases}
\end{figure}

\begin{figure}[H]
    \centering
    \caption{The evolution of the level values for infection and death rates and effective reproduction rate, $\beta_{l,t}$, $\nu_{l,t}$, and $eR_t$, over the sample from March 2020 until March 2022}
\begin{tabular}{cc}
\multicolumn{2}{c}{$\beta_{l,t}$} \\[-0.2em]
   \includegraphics[trim = 4.5cm 9cm 4cm 8.5cm, clip,width=0.35\textwidth]{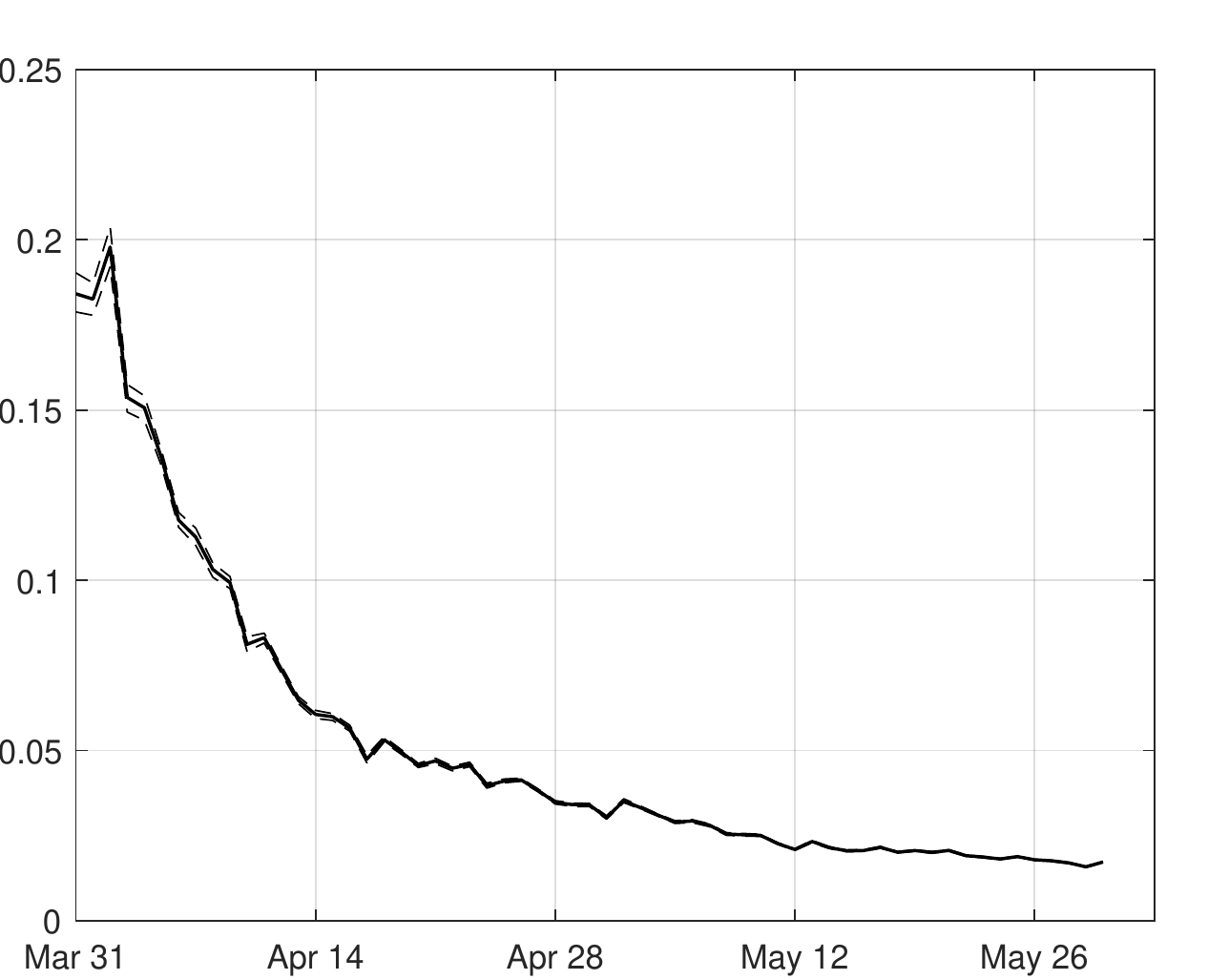}  
   &  \includegraphics[trim = 1.0cm 9cm 0cm 8.5cm , clip, width=0.55\textwidth]{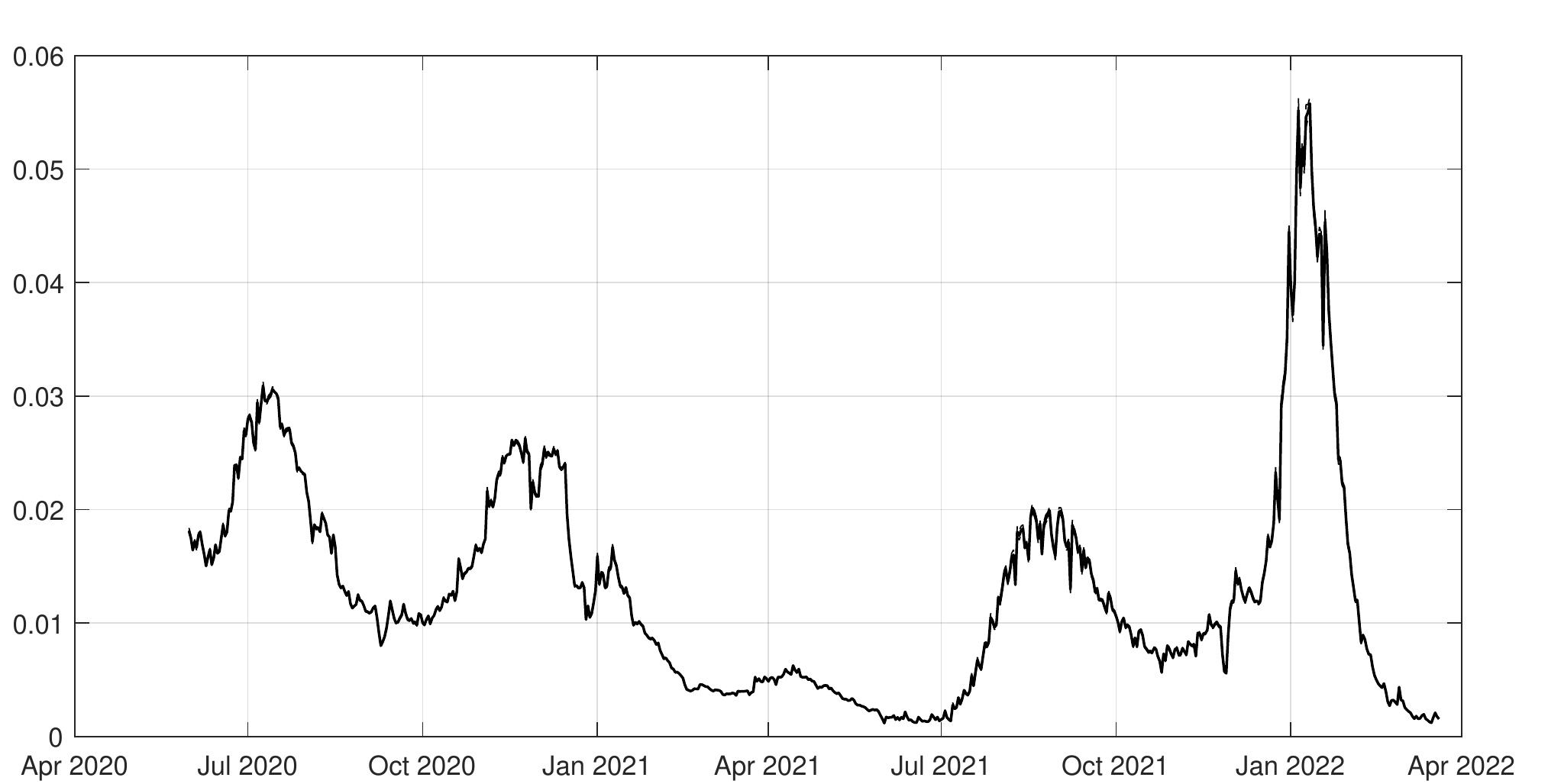}   \\ % \addlinespace
  \multicolumn{2}{c}{$\nu_{l,t}$} \\[-0.2em]
  \includegraphics[trim = 4.5cm 9cm 4cm 8.5cm, clip,width=0.35\textwidth]{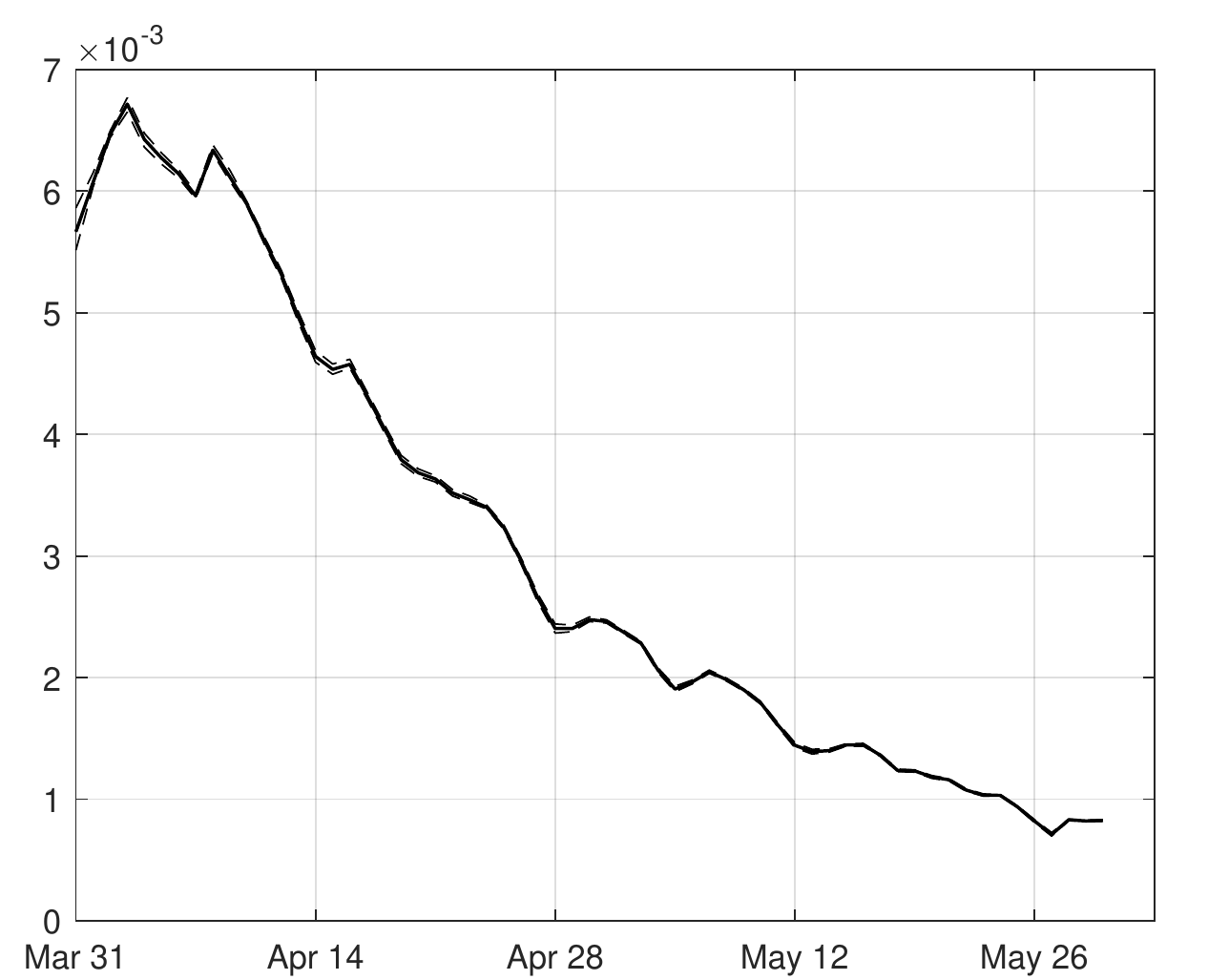}  
  &  \includegraphics[trim = 1.0cm 9cm 0cm 8.5cm , clip, width=0.55\textwidth]{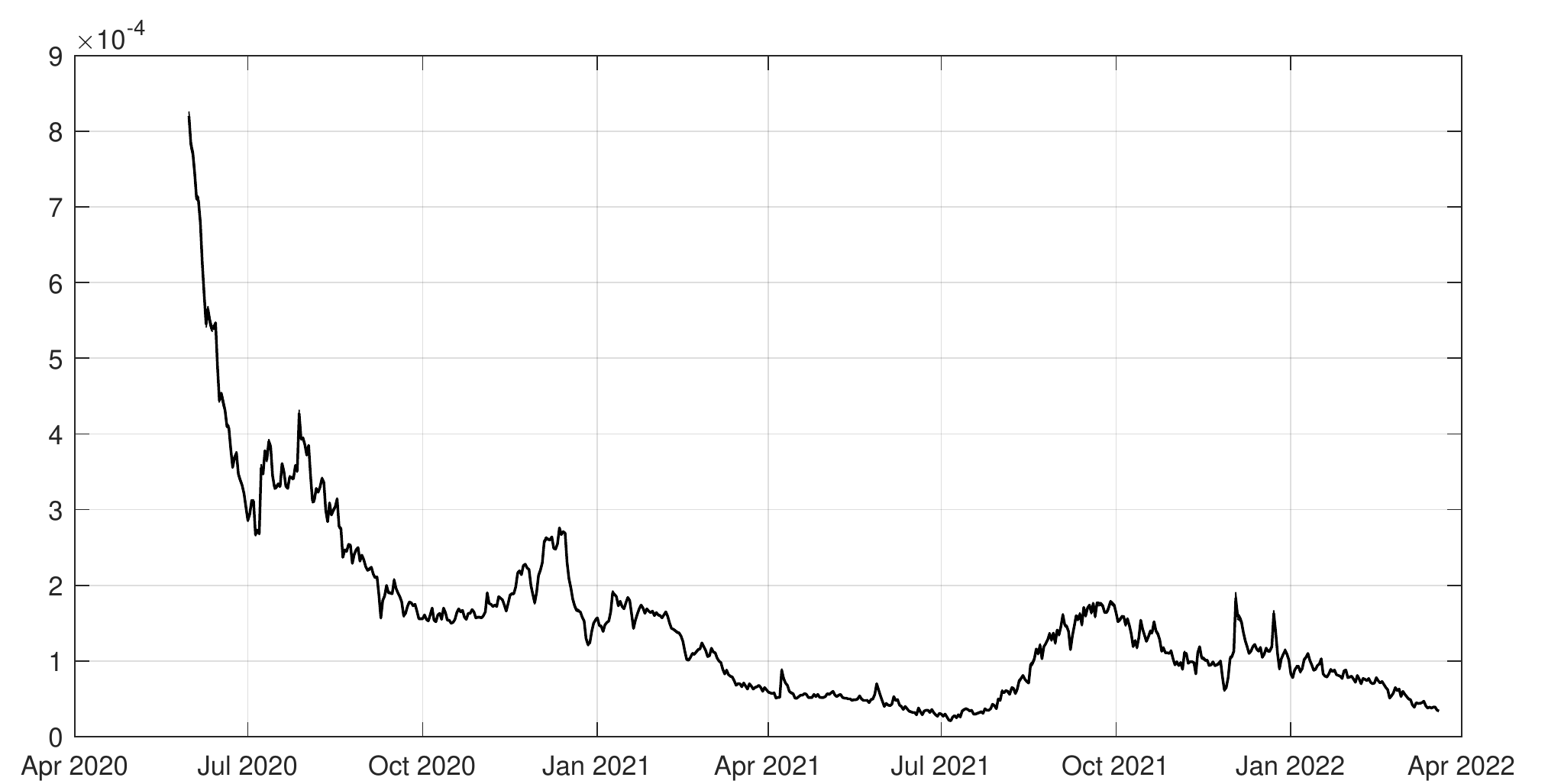} \\  %\addlinespace
  \multicolumn{2}{c}{$eR_{l,t}$} \\[-0.2em]
  \includegraphics[trim = 4.5cm 9cm 4cm 8.5cm, clip,width=0.35\textwidth]{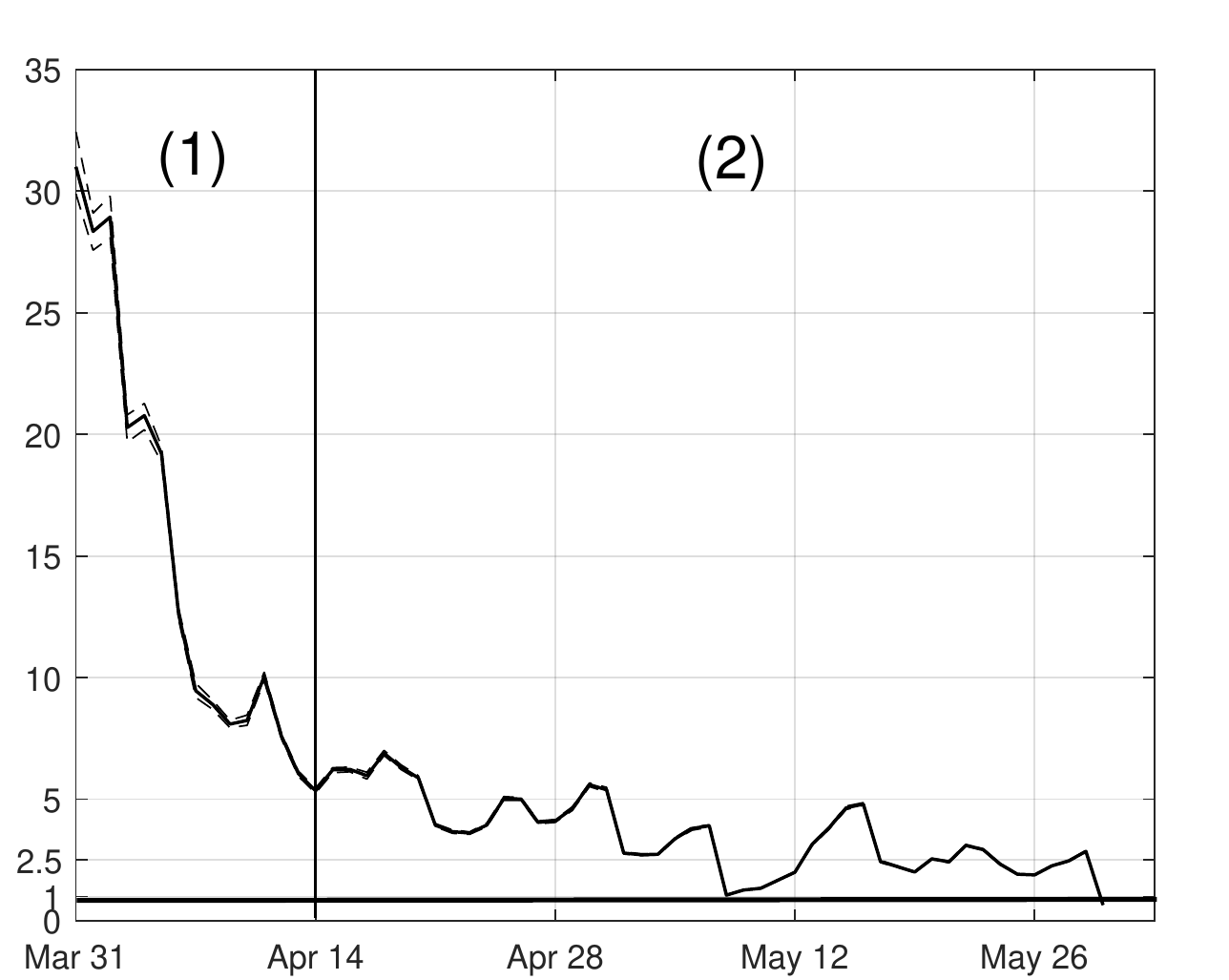}  
  &  \includegraphics[trim = 1.0cm 9cm 0cm 8.5cm , clip, width=0.55\textwidth]{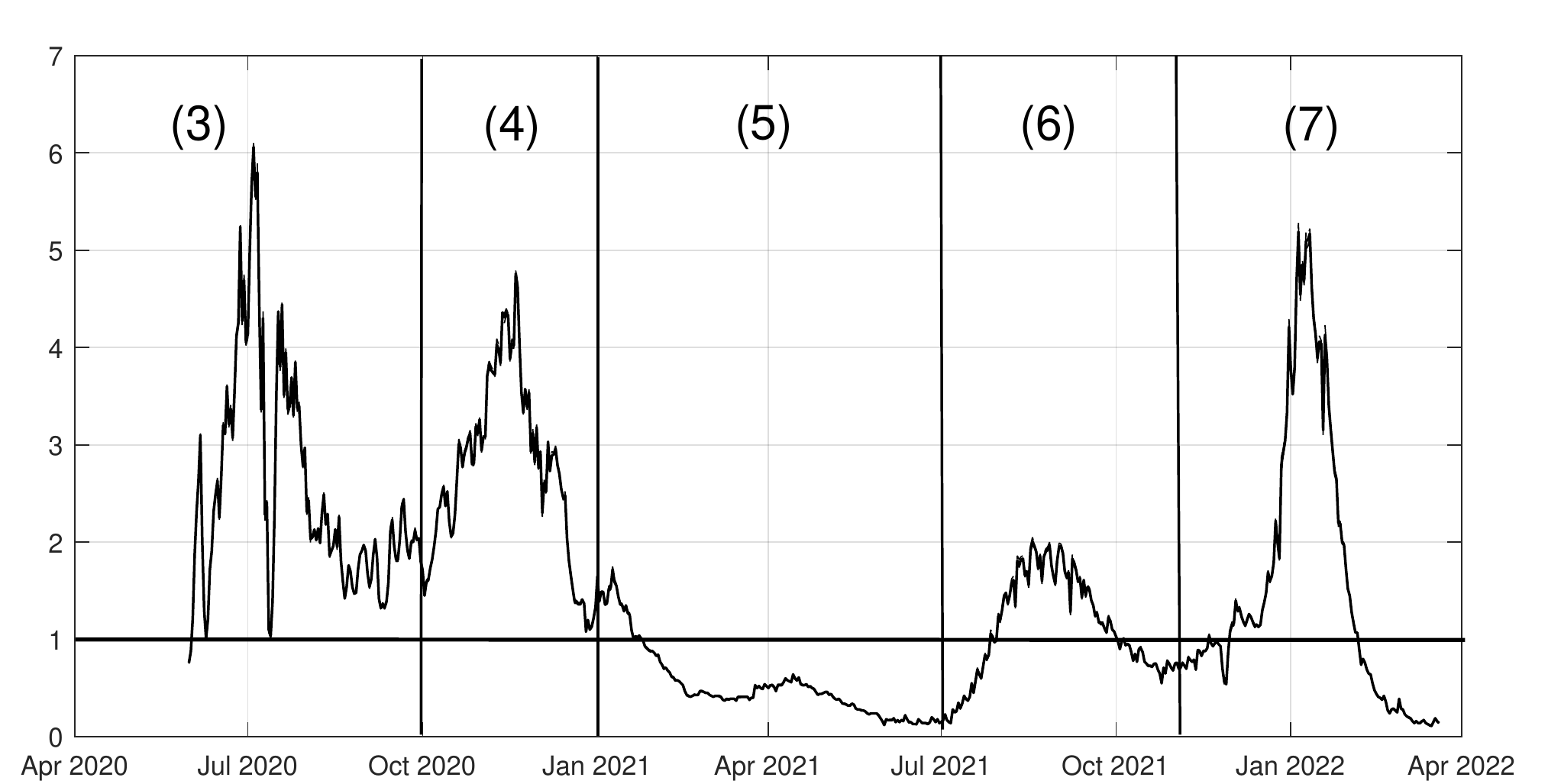} 
   \end{tabular}
    \label{fig:TVPSIRD_param}
\end{figure}

The effect of pharmaceutical or nonpharmaceutical interventions is reflected in the reproduction rate through the 'implied' model parameters as discussed in Section~\ref{subsec:Motivation}. These effects are all reflected as implied time variation in these parameters captured by the TVP-SIRD model, which we discuss in the remainder of this section\footnote{We also perform a statistical evaluation for the presence of time variation. In all the tests, the null hypothesis boils down to the hypothesis that the coefficients of the score functions are zero. As indicated by \cite{Calvori-etal-2014}, the test against the time-varying parameter alternative checks whether there is any autocorrelation in the scores of the fixed parameter model. If that is the case, such autocorrelations can be exploited to improve the model's fit by using the likelihood scores as drivers for the time-varying parameter as in the TVP-SIRD model. Our test results indicate decisively favor time variation in the model parameters. We thank an anonymous referee for pointing out this.}. Before discussing the evolution of the model parameters, we first start with displaying the fitted values of the daily number of confirmed and death cases in Figure~\ref{fig:FittedConfirmedDeathCases} to consider the overall performance of the TVP-SIRD model in fitting the pandemic dataset. 
\begin{center}
[Insert Figure~\ref{fig:FittedConfirmedDeathCases} about here]
\end{center} 
The left panel of the Figure~\ref{fig:FittedConfirmedDeathCases} shows the satisfactory fitting performance of the TVP-SIRD model for the daily confirmed cases. Both level and seasonal patterns could be matched using the model framework that can capture daily seasonal behavior in addition to level. We display the fitted values of the daily number of death cases on the right panel of Figure~\ref{fig:FittedConfirmedDeathCases}. This panel confirms the model's ability to capture daily death cases' level and seasonal patterns. Next, we display the evolution of the (level of the) model parameters and the estimated effective reproduction rate, $eR_t$, using the TVP-SIRD model in Figure~\ref{fig:TVPSIRD_param}.       
\begin{center}
[Insert Figure~\ref{fig:TVPSIRD_param} about here]
\end{center} 
In the first two graphs, we display the variation in the infection and death rates, and in the bottom set, we display the effective reproduction rate, $eR_t$.\footnote{The data on recovery ceased to be published after December 2020. We treat these periods of 2021 and 2022, where the recovery data is unavailable, as missing data. The estimate of the recovery rate remains constant for these periods when the data is missing since the score functions for these periods are set to 0; see \cite{creal2014observation} and \cite{lucas2016accounting} for a similar approach. Accordingly, we do not display this parameter's evolution, estimated as 0.0091, in large part of our sample period.} For clarity of demonstration, we display the evolution of the parameters and the effective reproduction rate in two subfigures representing two subperiods. On the left, we only display the periods until June 2020; on the right, we display the remaining periods. In these two subperiods, the scale of the variation of the parameters differs considerably, and providing two graphs for two different subperiods enhances the display's visual quality. Finally, in the bottom graph of the effective reproduction rate, we split the sample into seven enumerated subperiods with distinct characteristics to motivate the variation in the $eR_t$.

The first period labeled as (1), starting from early 2020 until April 2020, is the emergence period of the pandemic. During these periods, the World Health Organisation (WHO) officially declared the pandemic on March 11, and the national public health agency in the US imposed various measures to contain the pandemic, including the ban of large gatherings and travel restrictions. Until the end of April, 'stay-at-home' quarantines were effective in several locations, and many testing facilities for effectively isolating the infected cases were established. Our estimates suggest that the effective reproduction rate reduced from values as high as 35 to around 6 in mid-April. This reduction is in line with early studies reporting a basic reproduction rate (which is very close to the effective rate at the onset of the pandemic) of around 5.7 using the early dataset from Wuhan, China, the point of emergence of the pandemic, see \cite{sanche2020high} and around 5.3 for the US, see \cite{peirlinck2020outbreak}. In the second period (2), comprising the period from mid-April until June, the $eR_t$ steadily fluctuates around the value of 2.5 with the accumulation of the bulk datasets, similar to the studies reported elsewhere, see \cite{petersen2020comparing} for example for a comparison of SARS-COV-2 parameters to that of the SARS-COV and influenza viruses. The third period, (3), captures the summer period of 2020. This period experienced the economy-wide reopening and relaxation of various measures. Many large-scale events resulted in large gatherings\footnote{See for example New York Times article on 80$^{th}$ Motorcycle rally in South Dakota, where there were more than 400,000 audiences in the gathering, \url{https://www.nytimes.com/2020/11/06/us/sturgis-coronavirus-cases.html}.} that lead to an increase in the implied infection rate as can be seen in the first subfigure. The increase in the infection rate had overcome the increase in the death rate, which led the effective reproduction rate to increase to values around five again. The fourth period, (4), captures the winter period that includes the holiday season of Thanksgiving followed by the Christmas period, with an estimated number of more than 2 million people flying on airlines during the Thanksgiving period.\footnote{See the article \url{https://www.masslive.com/coronavirus/2020/11/thanksgiving-travel-many-americans/-flying-for-holiday-despite-cdcs-pleas-not-to-due-to-covid-19-transmission-risk.html} for example.} In addition to the changing mobility of the susceptible people, there was also a new variant of the virus, denoted as Alpha, first detected in December 2020. \cite{davies2021estimated} report that this variant is 43\%-90\% more transmissible than the predecessor lineage. Therefore, following the demonstration in subsection~\ref{subsec:Motivation}, the time-varying mixture of these two variants might be one underlying source of the increasing infection rate.  

The year 2021 started with a massive inoculation campaign with the availability of the Covid-19 vaccines with an efficacy rate as high as 90\%, see \cite{polack2020safety}. This period (5) experienced a significant and rapid drop in the effective reproduction rate, which fell below the critical value of 1 for the first time since the start of the pandemic. Therefore, in the first six months of 2021, the US successfully contained the pandemic, thanks to the vaccination campaign, which led to 67\% of the overall adult population receiving at least one dose. In the second half of 2021, the proportion of vaccinated people in the population has remained relatively steady, above 60\%. Furthermore, containment measures have also remained stable to a large extent. Therefore, the changes in the implied parameters in periods (6) and (7) mainly stem from the emergence of new variants with new structural parameter values. Indeed, in period (6), which spans the summer and early fall of 2021, the Delta variant was the dominant strain, according to the CDC estimates. The Delta variant seems to be around 60\% more transmissible than the already highly infectious Alpha variant, see \cite{callaway2021delta}, which can also be traced in the course of the infection rate and, thereby, the effective reproduction rate. Finally, in the last period, (7), which captures the remaining period until the end of March 2022, the Omicron variant has been the dominant variant which is much more contagious than the previous strains but less severe compared to those, see \cite{karim2021omicron}. This rapid infection rate surge due to the Omicron variant can be captured nicely using our model framework.
Moreover, the increase in the death rate remained relatively moderate and lower than that of the Delta variant, confirming the findings on the Omicron variant. As a result, the effective reproduction rate soared to as high as five. As of March 2022, the rate again fell below the critical value of 1. The full-sample findings demonstrate that the model with 'implied' time-varying parameters can capture various factors, such as changes in the number of susceptible people either due to pharmaceutical or nonpharmaceutical interventions or the emergence of new variants of the virus accurately and promptly, confirming stylized facts.

\subsection{Accounting for unreported cases}
\begin{figure}[H]
    \centering
    \caption{The evolution of the ratio of total infections to the number of reported infected cases}\vspace{-0.5cm}
\begin{tabular}{cc}
  \includegraphics[trim = 0cm 9cm 0cm 9cm, clip,width=0.65\textwidth]{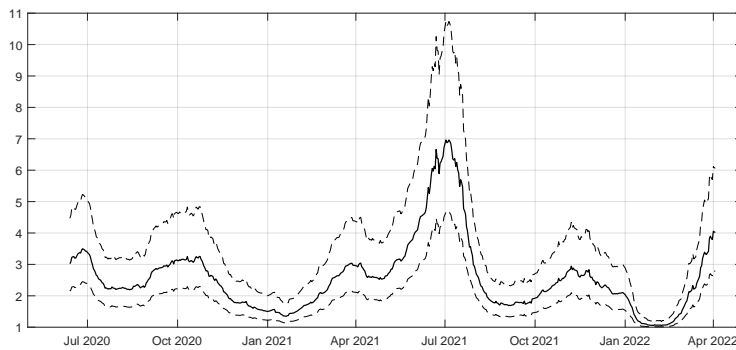}  
   \end{tabular}
    \label{fig:MF-TVPSIRD-delta}
\end{figure}

\begin{figure}[H]
    \centering
    \caption{The evolution of the level values for the death rate and effective reproduction rate, $\nu_{l,t}$ and $eR_{l,t}$ starting from March 2020 until March 2022}
\begin{tabular}{cc}
  \multicolumn{2}{c}{$\nu_{l,t}$} \\[-0.2em]
  \includegraphics[trim = 4.5cm 9cm 4cm 8.5cm, clip,width=0.35\textwidth]{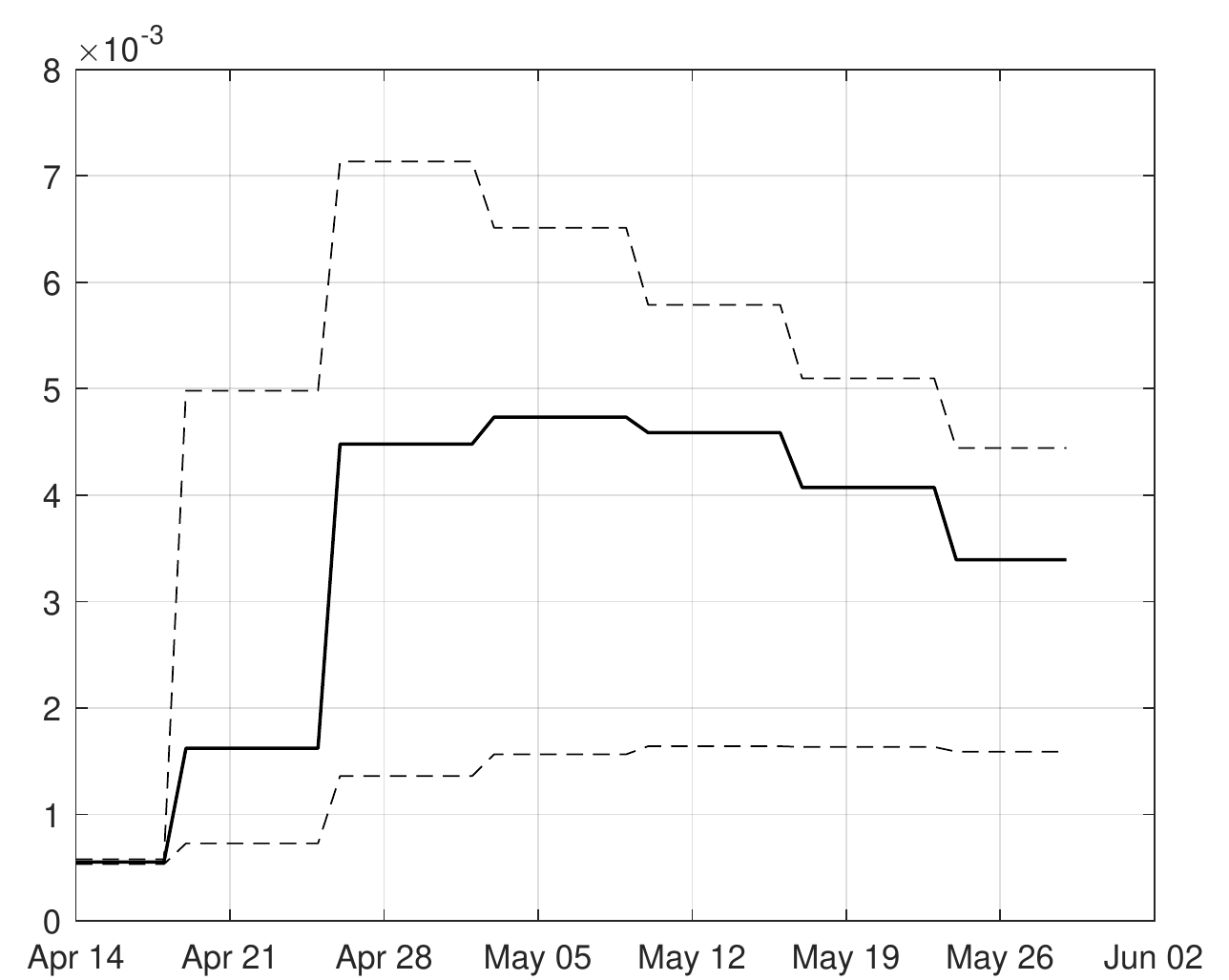}  
  &  \includegraphics[trim = 1.0cm 9cm 0cm 8.5cm , clip, width=0.55\textwidth]{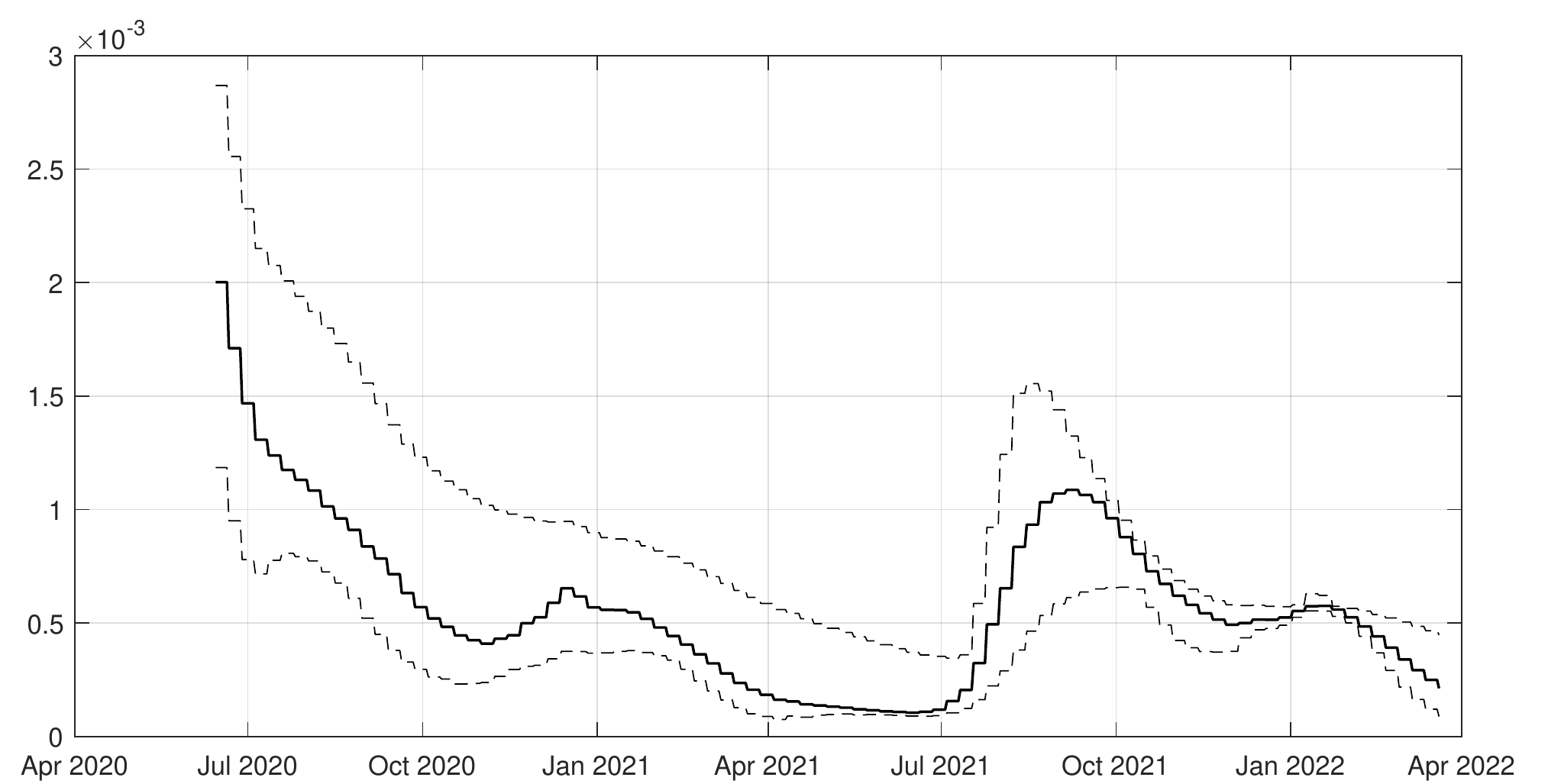}  \\ \addlinespace
  \multicolumn{2}{c}{$eR_{l,t}$} \\[-0.2em]
  \includegraphics[trim = 4.5cm 9cm 4cm 8.5cm, clip,width=0.35\textwidth]{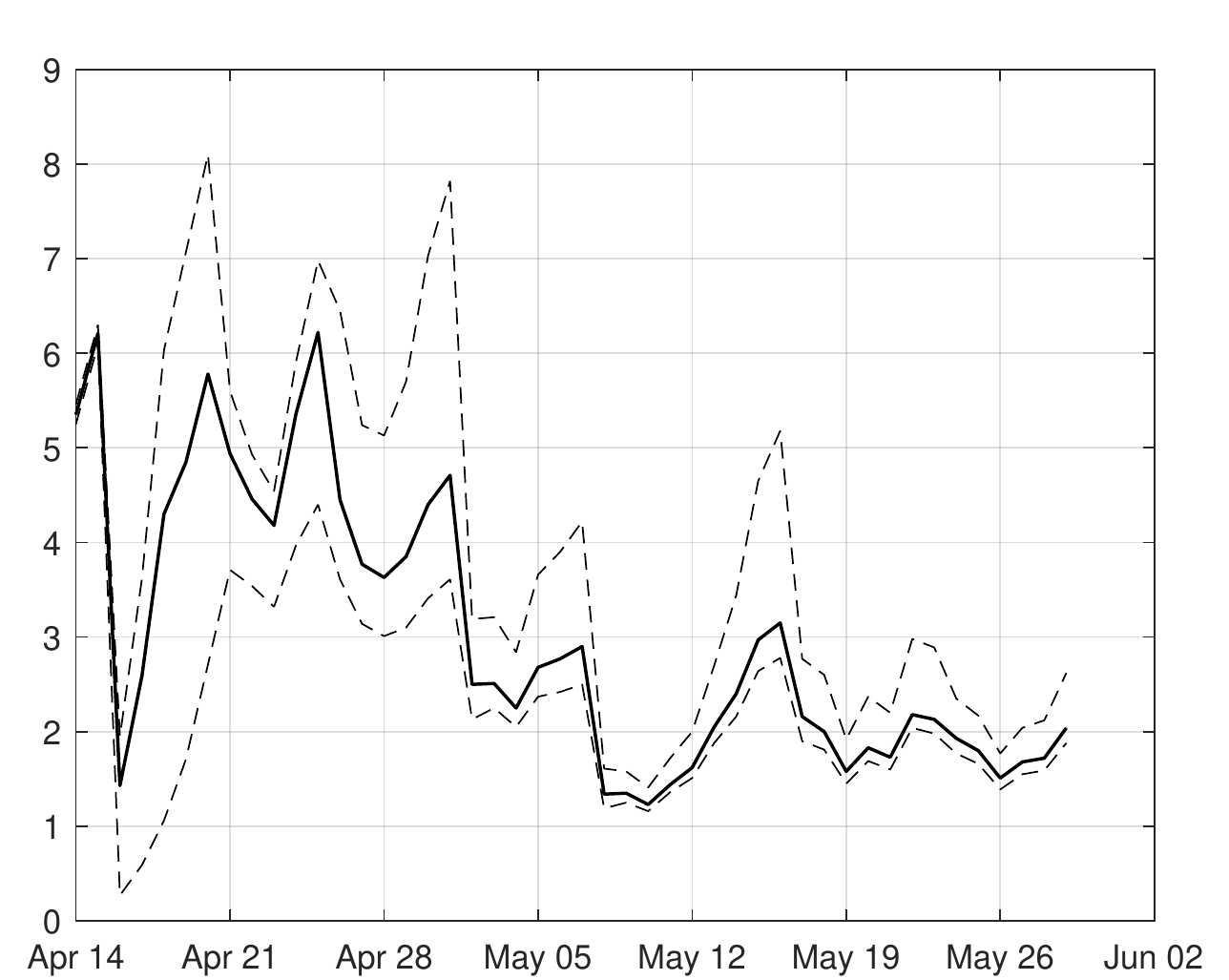}  
  &  \includegraphics[trim = 1.0cm 9cm 0cm 8.5cm , clip, width=0.55\textwidth]{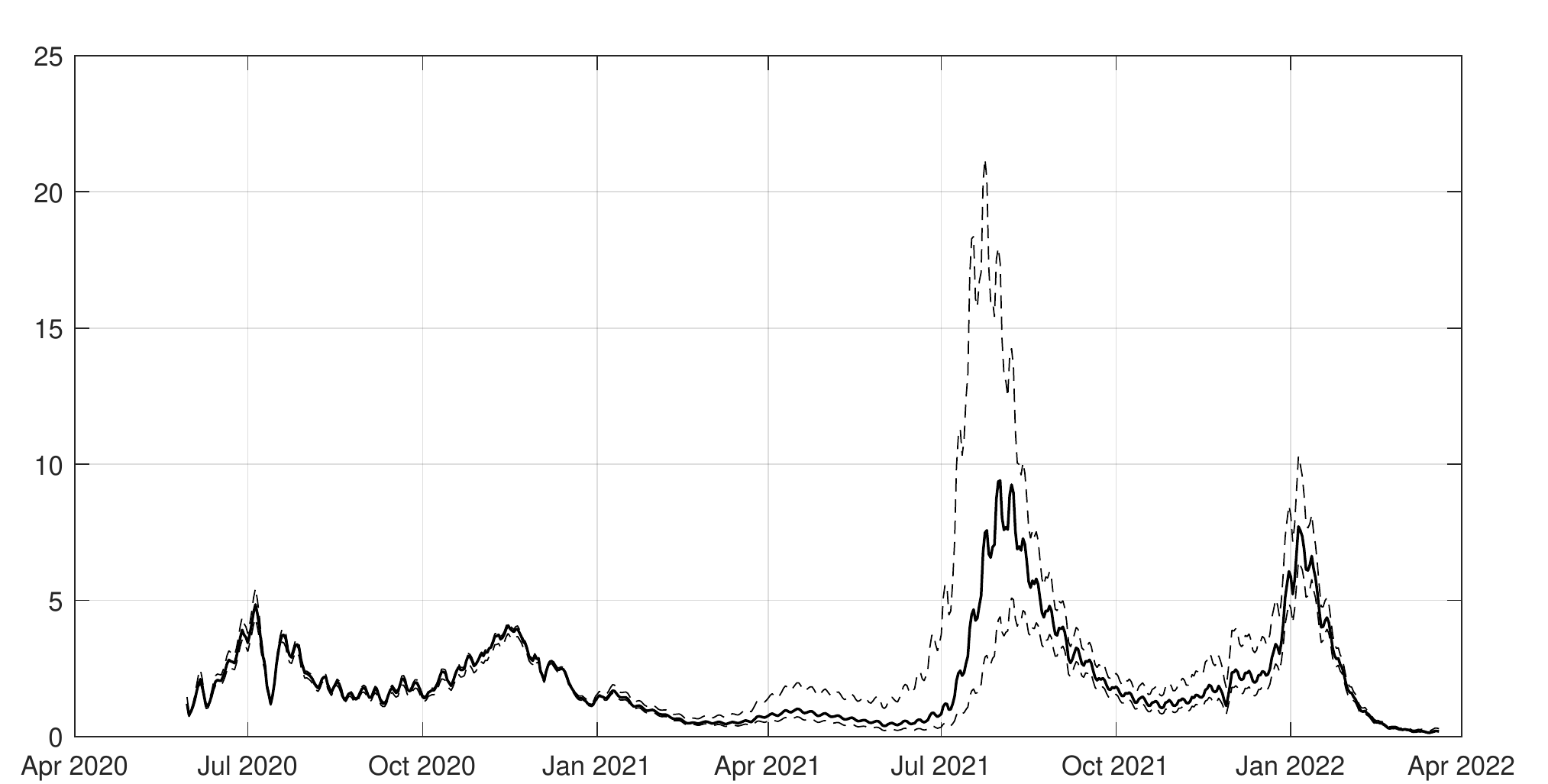} 
   \end{tabular}
    \label{fig:MF-TVPSIRD-param}
\end{figure}

The results discussed in previous sections are computed using official statistics, including only the reported cases. In this section, we present our findings when we account for this selection bias using the information on the number of excess deaths and the number of tests together with these tests' positivity rates. We display the evolution of the estimated rate of total infections to the number of reported infected cases, $1/(1-\delta_t)$, in Figure~\ref{fig:MF-TVPSIRD-delta}.
\begin{center}
[Insert Figure~\ref{fig:MF-TVPSIRD-delta} about here] 
\end{center} 
 Figure~\ref{fig:MF-TVPSIRD-delta} indicates that the actual number of infected cases, including the asymptomatic cases, might be three times more than the reported cases, especially during the peaks of the first two pandemic waves. However, our estimation results suggest a considerable uncertainty around this ratio with a 95\% credible interval between almost two and five on average. This finding is consistent with the CDC estimates reported as around 4 for the period until the end of 2020; see \cite{Reese_etal_2020} and the related web source\footnote{\url{https://www.cdc.gov/coronavirus/2019-ncov/cases-updates/burden.html}} for details. Similar findings are also reported by \cite{angulo2021estimation} based on the data from four regional and one nationwide seroprevalence surveys.\footnote{A seroprevalence survey uses antibody tests to estimate the percentage of people in a population who have antibodies against the virus. The number of people in a specific population who have been previously infected with the virus is estimated using these test outputs.} These serosurveys serve as a crucial data source for measuring the number of infected cases because the survey participants are selected randomly, thereby overcoming selection bias. Our results align with the reported results in \cite{angulo2021estimation}, where they estimate this rate as four using the nationwide serosurvey conducted during July-August 2020 in 47 states. While this is the case for most waves, including Delta and Omicron waves, throughout 2021, an important finding is in the late summer of 2021. Our results show that the total number of infected cases might be as high as seven times (with a wide 95\% credibility interval between [4-10]) of those reported cases. This finding is because, during this period, we observe a rapid surge in the number of excess deaths and relatively greater test positivity ratios. 
 This implies that the low number of confirmed cases in the summer of 2021 in the US might be mainly due to these unreported cases. According to the CDS estimates based on recurring serosurveys, the seroprevalence estimates, that is, the percentage of people with antibodies against the virus, soars from 20\% in July to around 30\% in September, which is in line with our findings.\footnote{Notice that the CDS report involves seroprevalence of infection-induced antibodies (nucleocapsid antibody) which is distinct from the vaccination-induced antibody (spike antibody). Therefore, these estimates genuinely represent the total infections; see \cite{Jones-etal-2022-Sero} for example for details.} We display the evolution of the death rates (based on the total number of deaths, including the excess deaths) and the effective reproduction rate in Figure~\ref{fig:MF-TVPSIRD-param}.  
\begin{center}
[Insert Figure~\ref{fig:MF-TVPSIRD-param} about here] 
\end{center} 
Considering the death rate, the discrete evolution of this parameter is due to the weekly frequency of the excess death dataset. This parameter is updated only at the weekly frequency when the data is observed and remains fixed for the other periods. While the death rate exhibits a similar pattern, we can track the surge in the rate in late summer of 2021, in accord with the previous discussion on the total number of infected cases, thanks to the excess death data. This finding is also confirmed in the evolution of the effective reproduction rate level, $eR_{l,t}$, where the $eR_{l,t}$ is computed as high as nine during these periods.   

	\section{Real-time performance of the models}
	\label{sec:Real-TimePerformance}
The results in the previous section display our findings based on the estimates using the full sample dataset. These results indicate that our flexible modeling structure can accommodate various forms of parameter changes reflecting the pandemic's course. However, exploring the model's real-time performance would uncover whether this additional flexibility brought by the time-varying parameters could provide timely and accurate information on the pandemic's real-time stance. Therefore, in this section, we discuss the model parameters' estimation results in real-time using the model with fixed parameters and the model with time-varying parameters. We aim to provide a thorough real-time analysis in the sense that we make use of the complete vintage data publicly available at the time of the prediction. Given that the pandemic data were revised substantially at times, using vintage data provides the actual predictive performance of the complete models. The vintage dataset is obtained from the Covid-19 Data Hub\footnote{https://covid19datahub.io/}, see \cite{guidotti2020} and \cite{guidotti2022} for details on the dataset. 

\subsection{Predictive performance at the daily frequency}
We use a rolling window for performing the SIRD model's estimations with fixed parameters rather than expanding window\footnote{We include the forecasting performance using an expanding window in the earlier versions of this paper. The results are decisively inferior compared to all moving window approaches. Therefore, we do not display those results here, but results are available upon request.} following the evidence of time variation in parameters in the previous section. Specifically, using the dataset from $t-M, t-M+1, \dots, t$, we estimate the SIRD model, and the resulting parameter estimates are those for the period $t$. We repeat this process by recursively adding one observation (and dropping one observation at the beginning of the sample for the rolling window). We consider three cases by setting $M=30$, $45$, and $60$, i.e., starting from one month of data up to two months of data. For capturing seasonality in these rolling window regressions, we consider daily dummy variables representing the days of the week. These models are denoted as RW-30, RW-45, and RW-60, respectively. For the TVP-SIRD model, we use the data up to period $t$ using an expanding window rather than a rolling window, as the parameters, in this case, are time-varying. We also include a restricted version of the TVP-SIRD model, where we allow for time variation only in the infection rate, $\beta$, denoted as TVP-SIRD-$\beta$, following similar approaches, see \cite{fernandez2020estimating} for example. Finally, we also include a time-varying parameter model that falls into the parameter-driven model category. In that case, we consider the computationally least costly alternative by imposing Normal distributions for observation and parameter evolution. For capturing the seasonality, we use the same model framework that we employ in the TVP-SIRD model with the critical distinction of including the error term in the state equations capturing seasonal patterns. The resulting specification leads to a standard inference using the Kalman filter and simulation smoother, which is still tractable in cases when the data is scarce; see \cite{durbin2012time} for details. This model is denoted as KF. 
For a given model, the predictive distribution of the observation at $t_{0}+1$ conditional on the information available at $t_0$, $\Omega^{t_{0}}$, is given by %\vspace{-0.4cm}
\begin{equation} 
p(y_{t_{0}+1}|\Omega^{t_{0}}) = \int f(y_{t_{0}+1}| \Phi) f(\Phi|y^{t_{0}} ) d \Phi, 
%\vspace{-0.2cm}
 \end{equation}
where $ f(\Phi|y^{t_{0}} )$ is the posterior distribution of the model parameters, estimated using the data until $t_0$, gathered in the parameter set, $\Phi$, given the observations until $t_{0}$. $p(y_{t_{0}+1}| \Phi)$ is the density of the observation $y_{t_{0}+1}$, which can be written as \vspace{-0.2cm}
\begin{equation}
\label{eq:PredDens}
f(y_{t_{0}+1}| \Phi) = \int \limits_{\theta_{t_0 +1}} f(y_{t_{0}+1}| \theta_{t_0 +1},\Phi) f(\theta_{t_0+1 } | \Phi,  \Omega^{t_{0}}).\vspace{-0.2cm}
\end{equation}
We can use the posterior simulator to obtain the distribution of the model parameters and estimate the predictive distribution using the draws from the simulator as $(y^{(m)}_{t_{0}+1}|\Omega^{t_{0}}, \Phi^{(m)})$, where $m$ represents the $m^{th}$ draw from the posterior simulator. We display the results involving Root Mean Squared Forecast Errors (RMSFEs) of the competing models relative to the TVP-SIRD model considering the prediction of the daily confirmed cases in Table~\ref{tab:OOS-C}. 
\begin{center}
[Insert Table~\ref{tab:OOS-C} about here]
\end{center} 
We perform equal predictive accuracy tests for the out-of-sample comparisons to evaluate the relative model performance. Specifically, for all the comparisons, we perform Diebold-Mariano (DM) type of pairwise comparison tests of equal predictive accuracy between the competing models with HAC standard errors and small sample correction suggested by \cite{harvey1997testing} using squared error contributions as loss functions. The cells with white backgrounds contain statistically insignificant values at the conventional significance level of 5\%. Table~\ref{tab:OOS-C} indicates a clear-cut result. The TVP-SIRD model outperforms all competing models up to 15 days horizon. This indicates the superior performance of our flexible modeling structure in the short and medium-term forecasting of the confirmed cases up to two weeks. This outperformance deteriorates for the horizons exceeding two weeks. In this case, although some models provide relative RMSFEs lower than unity, this relative performance is statistically insignificant at conventional significance levels. This is due to increasing uncertainty surrounding these point predictions leading to statistical insignificance. 

Focusing on pairwise evaluations, comparing the TVP-SIRD model with the TVP-SIRD-$\beta$ model reveals the importance of modeling time variation not only in the infection rate but also in the remaining parameters, at least for short and medium-term predictions. For the horizons longer than two weeks, the two models perform alike with relative RMSFEs very close to unity. Comparison of the TVP-SIRD model with the parameter-driven model with Normal distributions for the observables and the parameters, denoted as KF, indicates the merits of deterministic updating with a proper specification of the data structure over more flexibility of the parameter-driven models. In this case, the TVP-SIRD model performs significantly better than the KF model for up to 10 days, and the two perform statistically indifferent for longer horizons. Finally, the trade-off between the flexible and less flexible models is apparent when we compare short and long-horizon performances of the regressions with a 30-day moving window versus a 60-day moving window. While the regressions with a 30-day moving window outperform the counterpart with a 60-day moving window for the predictions up to two weeks due to flexibility, the latter model outperforms the former due to the reduction in the variance despite the increasing bias.  

We display the results involving RMSFEs of the competing models relative to the TVP-SIRD model when we consider the prediction of the daily death cases in Table~\ref{tab:OOS-D}. 
\begin{center}
[Insert Table~\ref{tab:OOS-D} about here]
\end{center} 
Table~\ref{tab:OOS-D} indicates mixing results. First, comparing the TVP-SIRD model with the TVP-SIRD-$\beta$ model indicates the importance of the time variation in the death rates. In this case, the TVP-SIRD model with the time-varying death rate outperforms the TVP-SIRD-$\beta$ with the fixed death rate at all horizons, and this outperformance is statistically significant. Comparison of the TVP-SIRD model with the KF model indicates that the former outperforms the latter significantly at all horizons except the 1-day ahead forecast, indicating the importance of proper modeling of pandemic count data. Finally, we compare the TVP-SIRD model performance with the regression models. The TVP-SIRD model performs better than the regression model with a 60-day moving window at all horizons except the 30 days horizon. When the window size is shortened to 45 and 30-day leading to more flexibly changing parameters, the superior performance of the TVP-SIRD model remains significant at longer horizons exceeding ten days. However, the predictions become statistically indifferent for short horizons thanks to the flexibility of these regression models with shorter windows. Overall, it seems that the flexibility of the TVP-SIRD model pays off even more at longer horizons for predicting the daily death cases compared to confirmed cases.\footnote{The predictive performance of the daily death cases is closely related to the number of ICU patients due to Covid-19, which is a critical factor for the decision-makers, with a correlation exceeding 0.8. The predictive results of forecasting the number of ICU patients are very similar to those of death cases, and these are presented in Section H of the supplementary material.}

\subsection{Predictive performance at weekly frequency}
In the previous section, we display a horse race for the predictive performances of a set of competing models at a daily frequency. However, since the outburst of the pandemic, many models, including various forms of epidemiological models, curve fitting frameworks, or machine learning setups, have predicted the pandemic's key variables, including confirmed and death cases. Luckily, the CDC-funded Influenza Forecasting Centers of Excellence worked closely with global, federal, state, and local public health officials to integrate infectious disease forecasting in a so-called forecast hub providing predictions of the outstanding forecasting sources.\footnote{See \url{https://covid19forecasthub.org/doc/. We thank an anonymous referee for pointing out this source.} for further details on this initiative.}. In this section, we compare the TVP-SIRD model with these prominent competitors. Since these forecasts are provided at weekly frequency, we estimate the TVP-SIRD model at the weekly frequency in real time using vintage data as in the previous case.\footnote{We also evaluate the weekly forecasts by aggregating our daily forecasts. However, this yields worse results compared to using weekly data for estimation. Therefore, we provide the forecasting results regarding weekly data setup.} We discard the forecasts that have less than 30 forecast readings, leaving us with 19 forecast sources for the prediction of the weekly confirmed cases and 28 for the prediction of the weekly death cases. We display the forecast sources that range from Microsoft to MIT-based models in the supplementary material in Table G.3.

We display the number of models that our TVP-SIRD model outperforms in Table~\ref{tab:ForecastHub} for horizons including $h=1,2,3,4$, i.e., from the 1-week horizon up to the 1-month horizon.  
\begin{center}
[Insert Table~\ref{tab:ForecastHub} about here]
\end{center}
Table~\ref{tab:ForecastHub} reveals that in the short horizons, the TVP-SIRD model can beat the majority of these outstanding forecast sources with 11 outperformance out of 19 sources for the prediction of the confirmed cases and 19 out of 28 sources for the prediction of the death cases considering 1-week horizon. However, this superior predictive ability monotonically erodes with the increasing forecast horizon. In line with the prediction results using daily frequency, the TVP-SIRD model is more successful in predicting the death cases at long horizons compared to the confirmed cases with better performance than 30\% (20\%) of the forecast sources at 3-week (4-week) horizon for death cases versus 15\% (10\%) for confirmed cases prediction. Therefore, we conclude that the TVP-SIRD model successfully predicts the critical Covid-19 pandemic-related variables at the short and medium horizon. It performs comparably to the leading pandemic forecasting tools at long horizons.   

\begin{figure}[H]
    \centering
    \caption{The evolution of the relative RMSFE of the ensemble model's 1-week ahead predictions relative to those of the TVP-SIRD model}
\begin{tabular}{cc}
 Weekly confirmed cases &  Weekly death cases \\[-0.2em]
 \includegraphics[trim = 1cm 9cm 1cm 9cm, clip,width=0.48\textwidth]{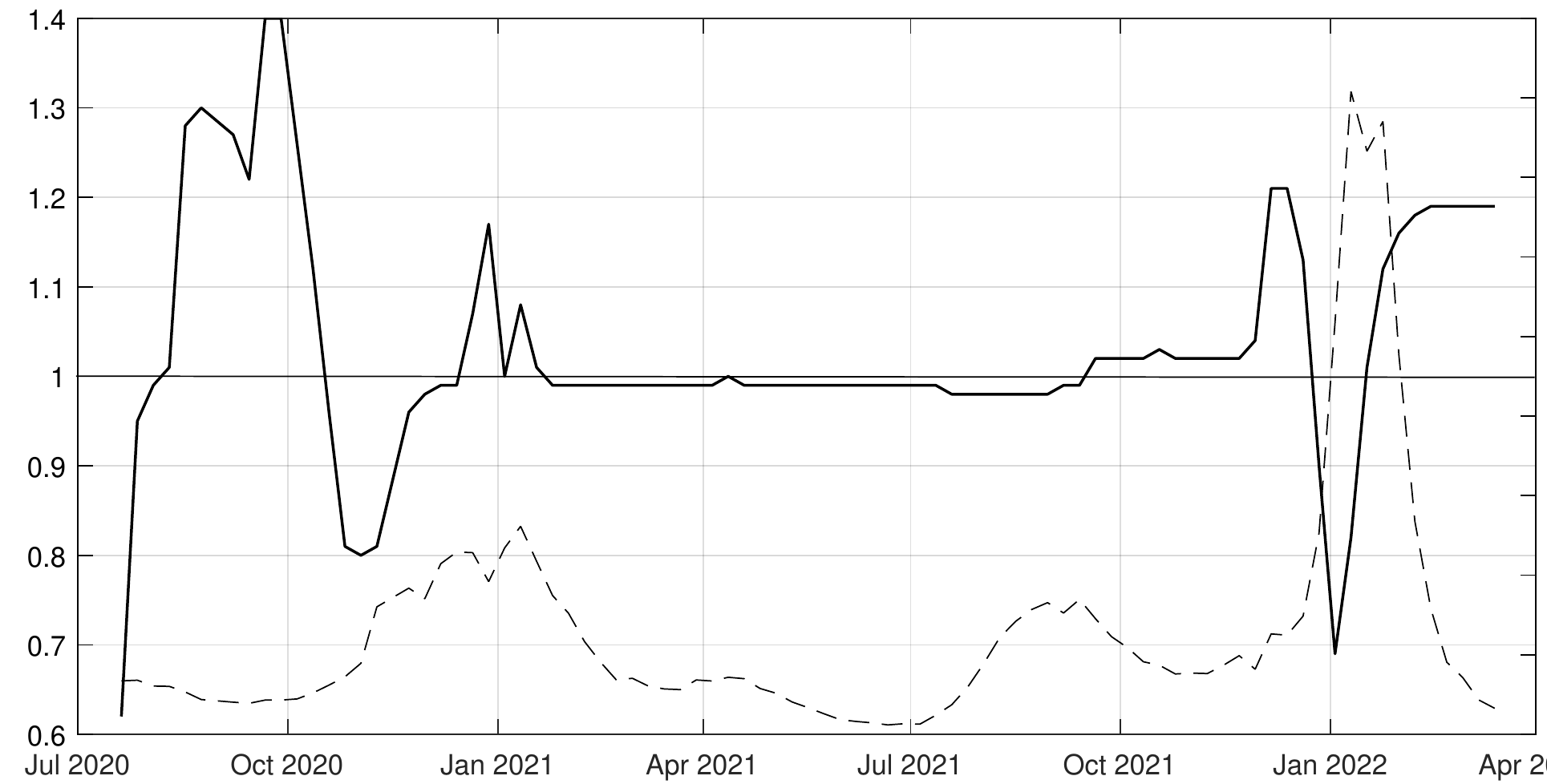} 
& \includegraphics[trim = 1cm 9cm 1cm 9cm, clip,width=0.48\textwidth]{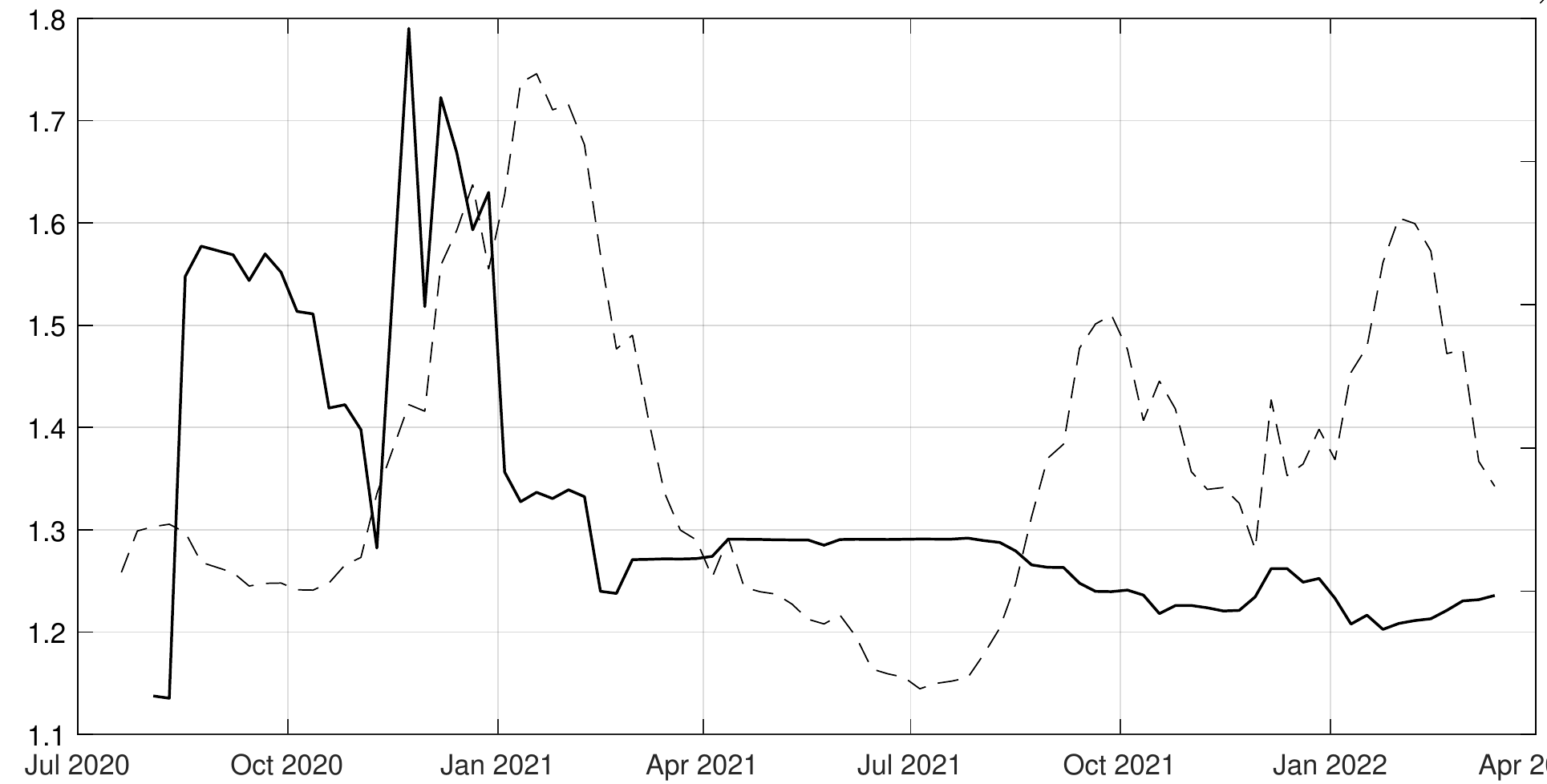}  \end{tabular}
    \label{fig:Ensemble1week}
\end{figure}

We provide a more detailed picture of the dynamic performance of the TVP-SIRD model over time compared to the forecasting tools in Figure~\ref{fig:Ensemble1week} for the 1-week horizon.\footnote{We display the results for longer horizons in Section G of the supplementary material.} 
\begin{center}
[Insert Figure~\ref{fig:Ensemble1week} about here]
\end{center}
Rather than providing pairwise comparisons with every single model, Figure~\ref{fig:Ensemble1week} displays the evolution of relative RMSFEs of the TVP-SIRD model with an ensemble model (EM) over time where relative RMSFEs (rRMSFE) are computed recursively in real time using vintage datasets. The EM is computed using a forecast combination scheme generated using the space of individual models. In line with the advantages of forecast combinations over individual models in many settings, the EM provides better predictions than the individual forecast sources. Thus, it is a gold standard in predicting confirmed and death cases; see \url{https://covid19forecasthub.org/doc/reports/} for details. In Figure~\ref{fig:Ensemble1week}, we also include the actual number of cases to compare the relative predictive performance taking the timing of the various phases of the pandemic into account. In the left and right panels of Figure~\ref{fig:Ensemble1week}, we display the RMSFE of the TVP-SIRD model relative to the EM for the prediction of the confirmed cases and death cases, respectively. A value lower than unity indicates the better performance of the TVP-SIRD model relative to EM. Considering the confirmed cases, on average, the TVP-SIRD model performs closely to the EM as the rRMSFE is very close to unity in most periods. A striking finding is that the TVP-SIRD model performs better than the EM, specifically at the onset of the pandemic waves. This result indicates that the TVP-SIRD model provides timely predictions at the onset of the pandemic waves reflecting the flexible model structure that can immediately accommodate the changing conditions. However, this picture reverses when the pandemic wave is at its peak. Once the data on the new pandemic wave is accumulated, the EM provides better predictions, especially on the timing of the turning point down the hill. Focusing on predicting weekly death cases at 1-week horizon, we observe that EM performs much better than the individual forecast sources. Unlike the previous comparison of the TVP-SIRD model to individual forecast sources, the EM model performs better than the TVP-SIRD model over time as relative RMSFEs exceed unity persistently. Still, the pattern discussed in the prediction of the confirmed cases is also apparent here. Again, the performance of the TVP-SIRD model improves at the onset of the pandemic waves and deteriorates once the pandemic's peak is passed. Overall, our results indicate that the TVP-SIRD model performs favorably well at daily and weekly frequency against very compelling competitors in forecasting of key Covid-19 pandemic aggregates.   

\section{Potential extensions}
\label{sec:Extension}
\begin{figure}[H]
    \centering
    \caption{The evolution of the number of active infected cases in countries used for factor TVP-SIRD model estimation}\vspace{-0.3cm}
\begin{tabular}{ccc}
Germany &  Italy &  Brazil\\[-0.3em]
 \includegraphics[trim = 10mm 90mm 10mm 85mm, clip, width=0.33\textwidth]{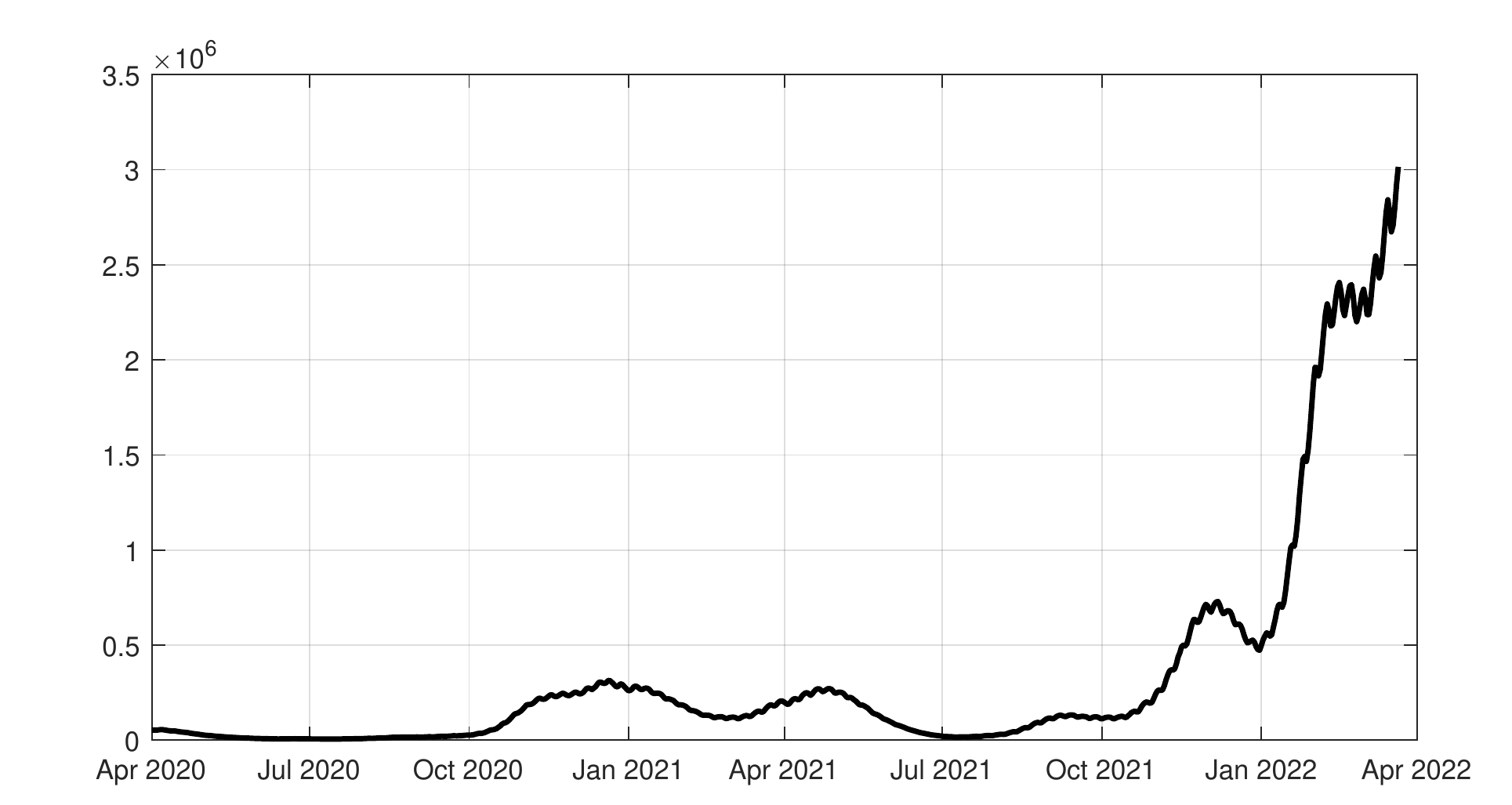} 
  &   \includegraphics[trim = 10mm 90mm 10mm 85mm, clip,width=0.33\textwidth]{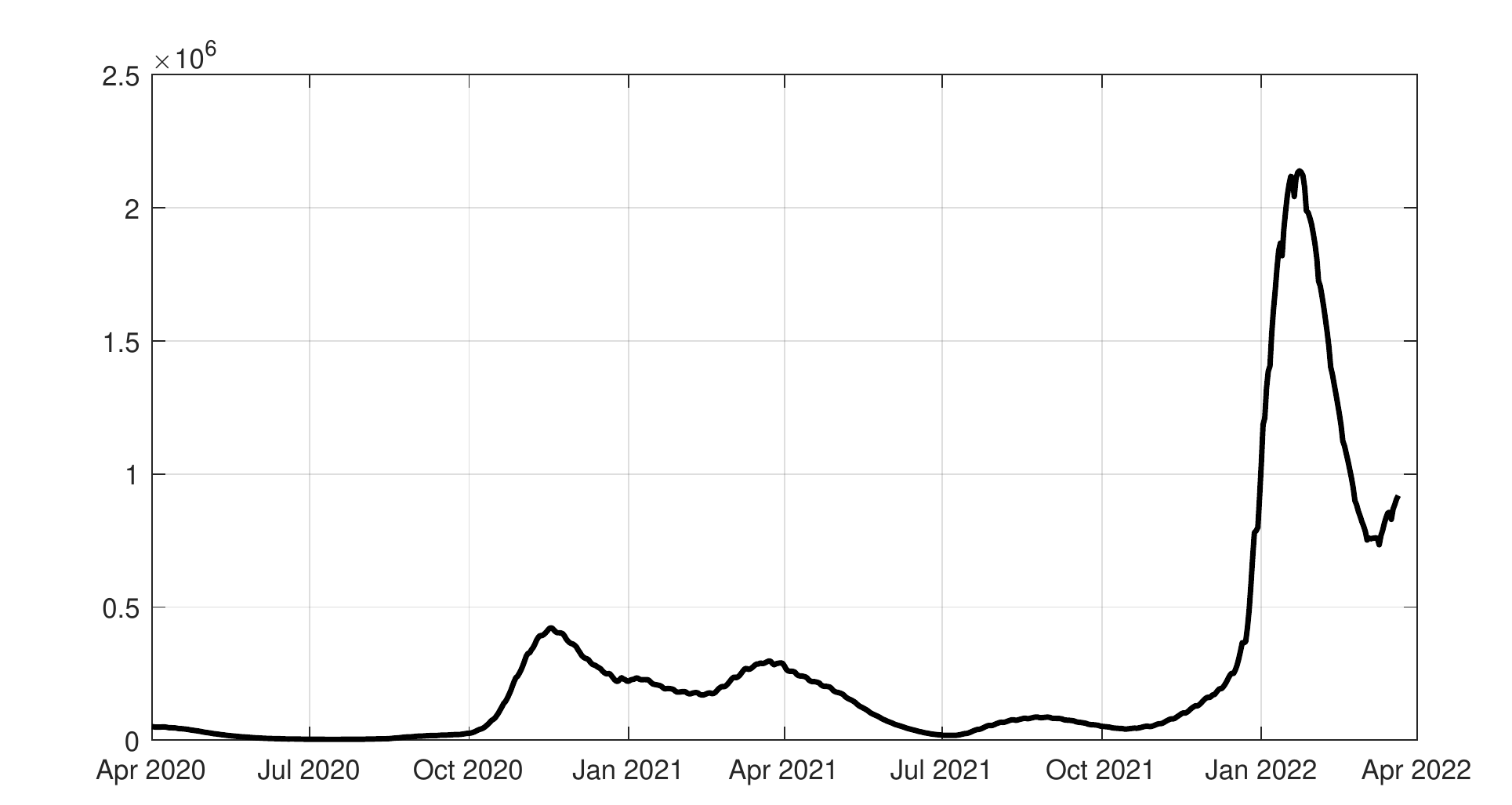}  
  &   \includegraphics[trim = 10mm 90mm 10mm 85mm, clip,width=0.33\textwidth]{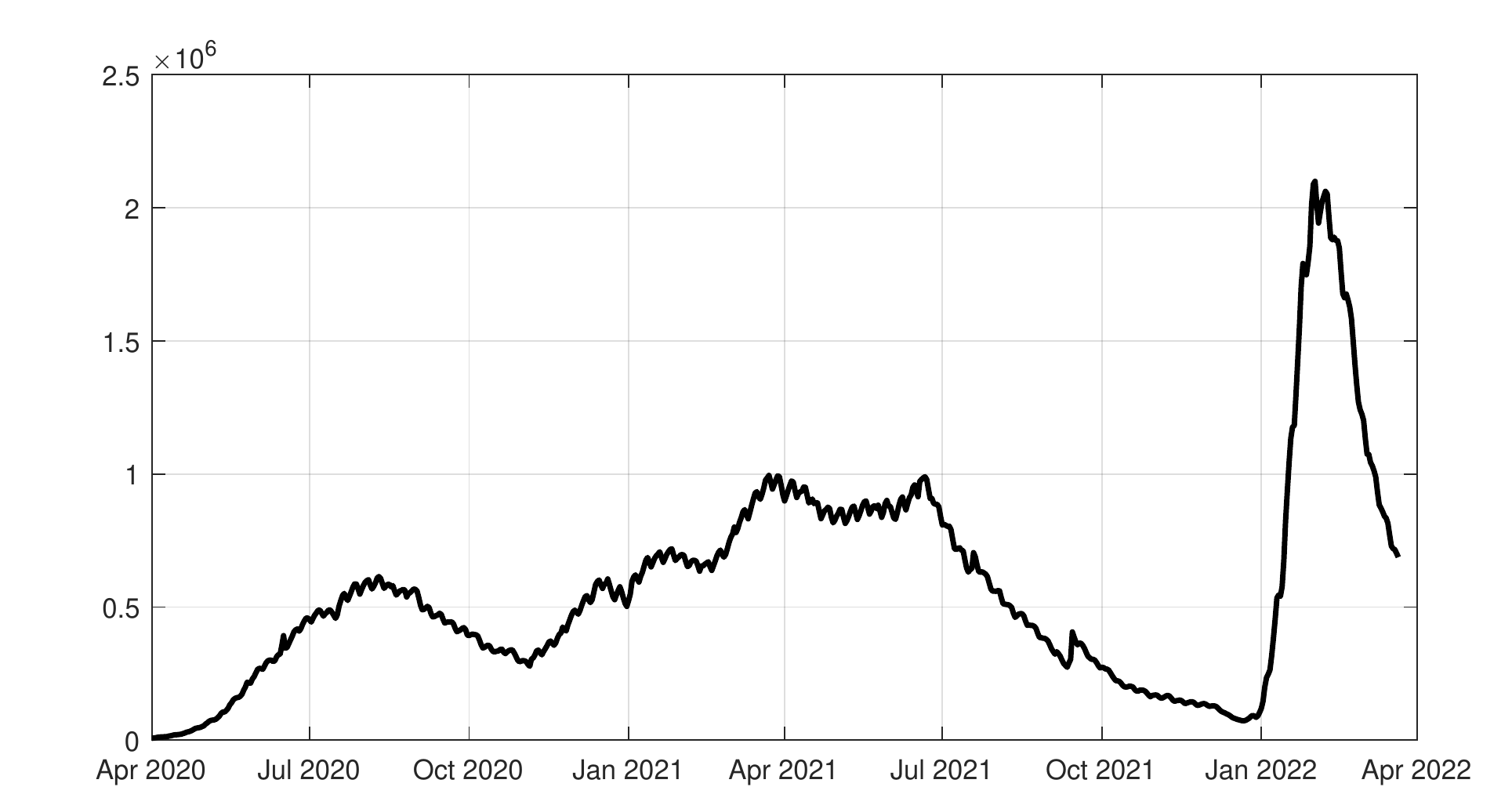}  
   \end{tabular}
    \label{fig:Factor-AIC}
\end{figure}

\begin{figure}[H]
    \centering
    \caption{The evolution of common infection rate and country-specific $eR_t$s}\vspace{-0.3cm}
\begin{tabular}{cc}
Infection rate factor & $eR_t$ of Germany \\[-0.4em]
     \includegraphics[trim = 10mm 90mm 10mm 85mm, clip,width=0.50\textwidth]{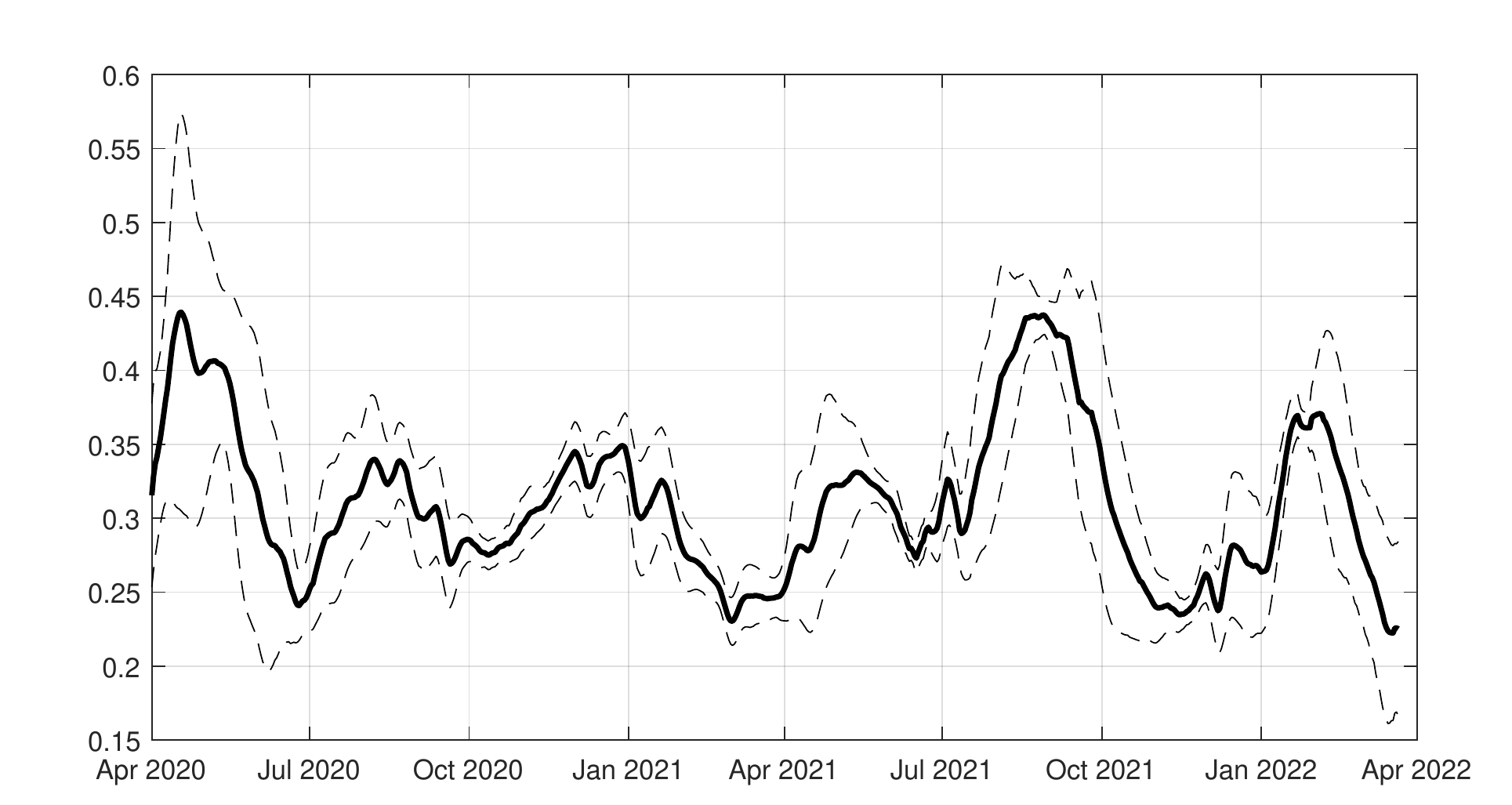}  
  &  \includegraphics[trim = 10mm 90mm 10mm 85mm, clip, width=0.50\textwidth]{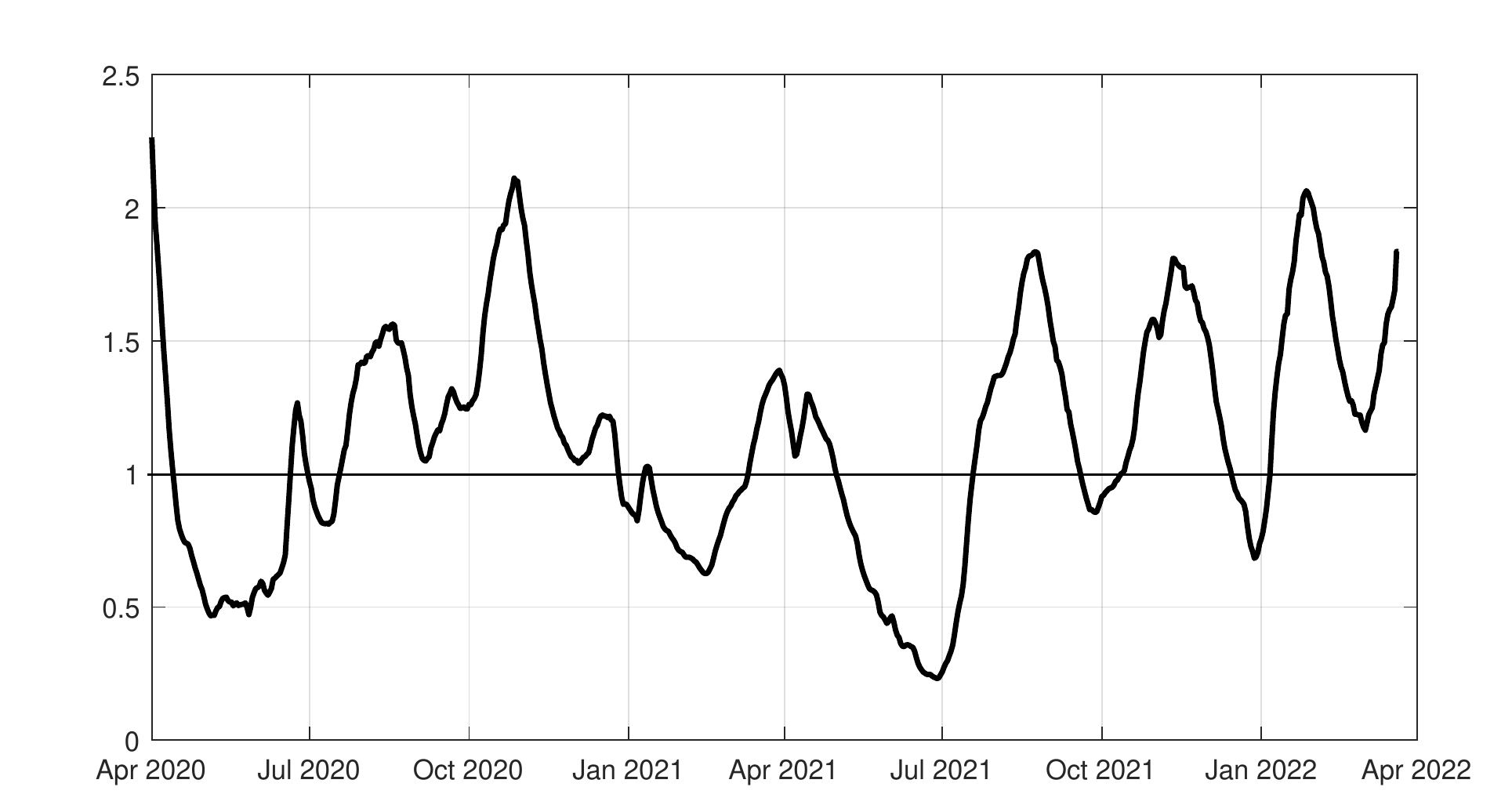} \\
 $eR_t$ of Italy&  $eR_t$ of Brazil\\[-0.4em]
     \includegraphics[trim = 10mm 90mm 10mm 85mm, clip,width=0.50\textwidth]{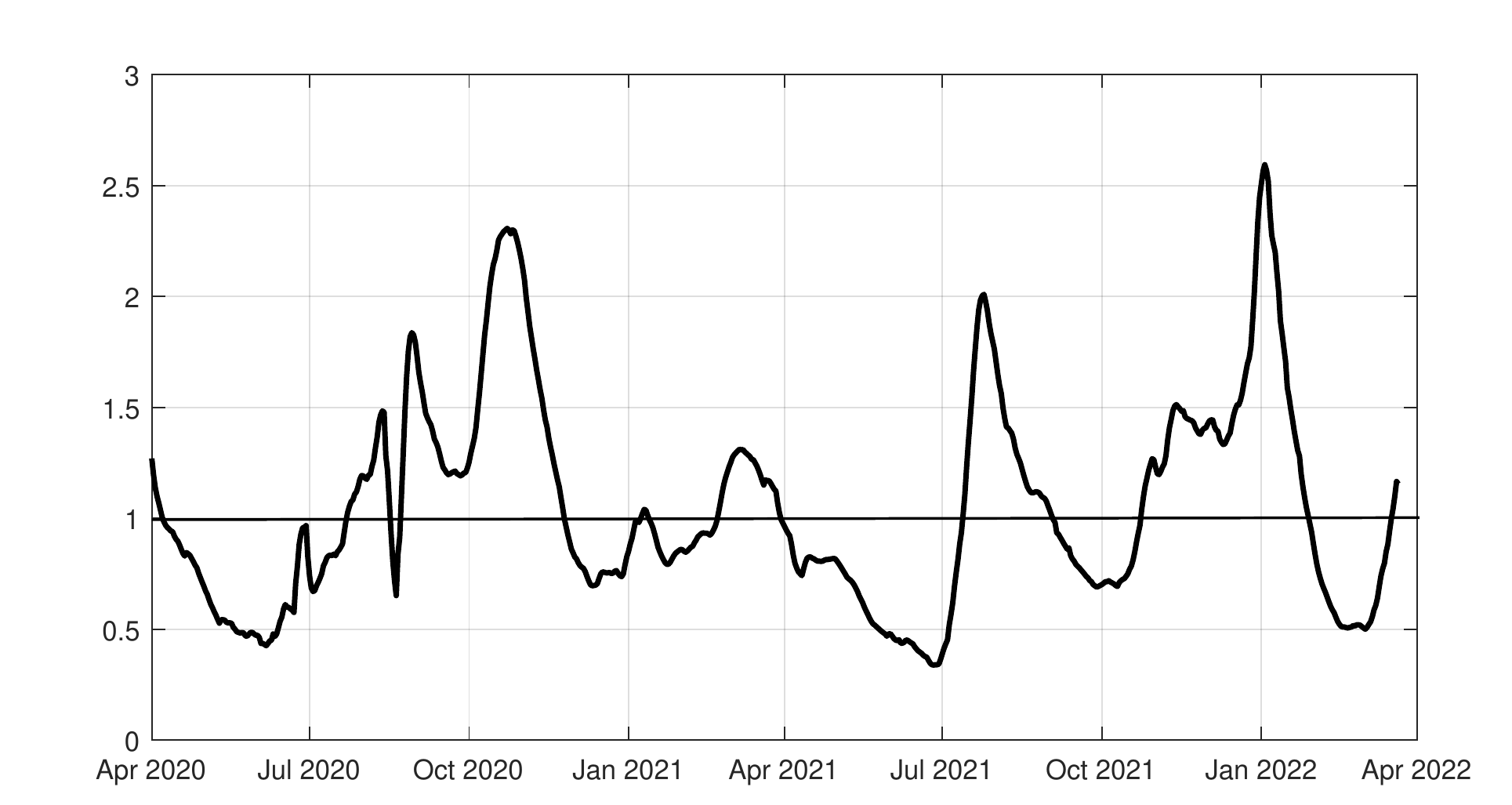}  
  &   \includegraphics[trim = 10mm 90mm 10mm 85mm, clip,width=0.50\textwidth]{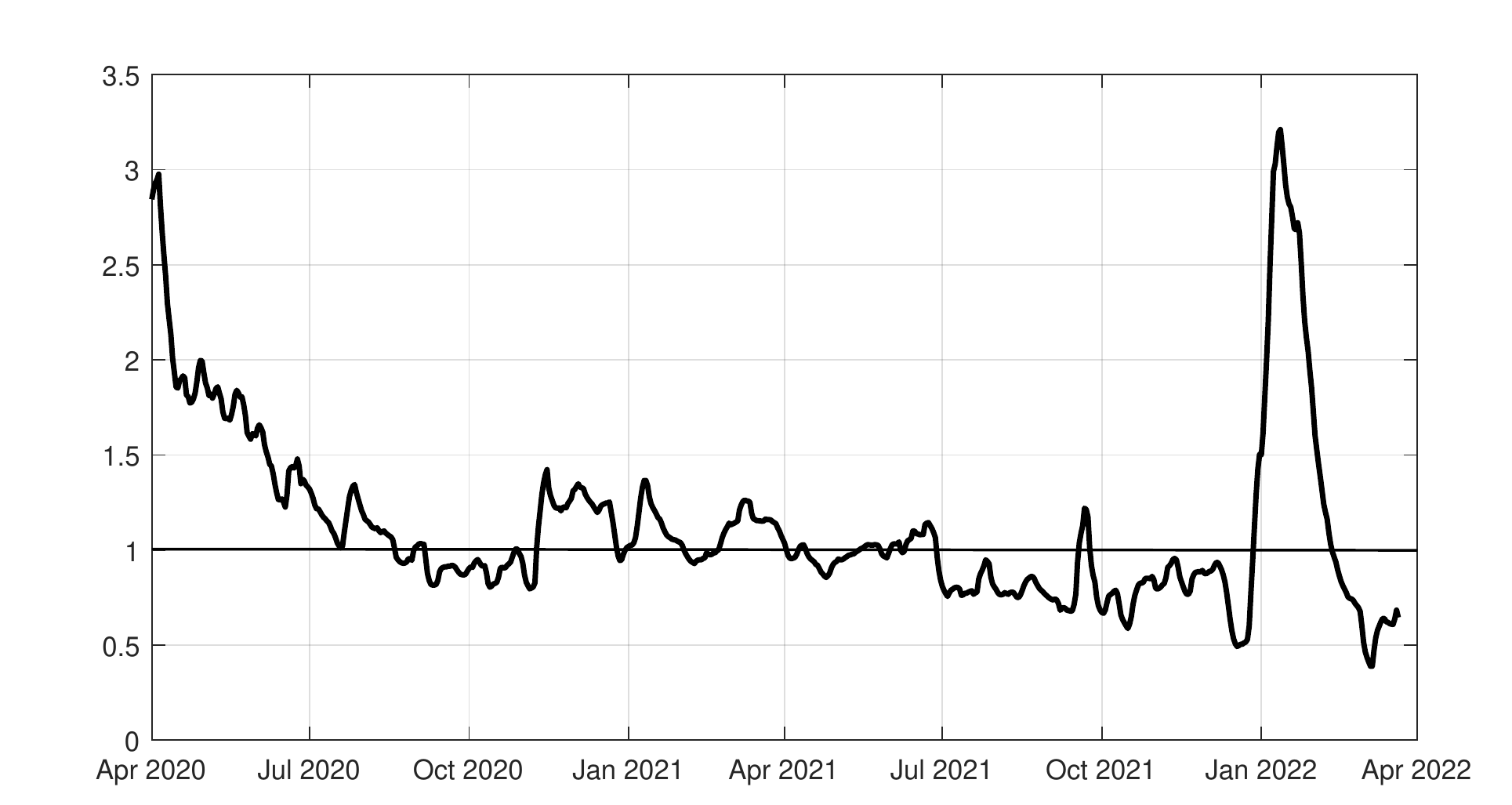}  
   \end{tabular}
    \label{fig:Factor-Output}
\end{figure}

In the previous sections, we display the potential of the TVP-SIRD model both in in-sample fit and out-of-sample forecasting. This section provides a potential extension to the TVP-SIRD model. Since the pandemic is a global phenomenon, multiple countries have had common experiences, with some countries having relatively larger part of idiosyncratic variations. Departing from this observation, we extend the model to a multi-country setting using a factor model structure. Consider the following model for country $i=1,\dots,K$,
\begin{equation}
\label{eq:Factor-TVP-SIRD}
    \begin{array}{rcl}
    \Delta C_{i,t}|\Omega_{t-1} &\sim&  Poisson( \beta_{i,t} \frac{S_{i,t-1}}{N_i}I_{i,t-1}) \\[-0.2em]
	\Delta Rc_{i,t}|\Omega_{t-1} &\sim&  Poisson( \gamma_{i,t} I_{i,t-1})    \\[-0.2em]
	\Delta D_{i,t}|\Omega_{t-1} &\sim&  Poisson( \nu_{i,t} I_{i,t-1})    \\[0.4em]
    \Delta I_{i,t}   &=&  \Delta C_{i,t} - \Delta Rc_{i,t} - \Delta D_{i,t}.\\
    \end{array}
\end{equation}
We assume that country-specific parameters of the TVP-SIRD model admit a factor structure for their level, while the seasonal patterns are idiosyncratic.\footnote{Imposing a factor structure to the seasonal component is a straightforward extension of the model. However, our experience with this model shows that identifying a common seasonal factor poses more challenges than a level factor. Since the level factor is the key component of the parameters, we consider a factor structure only in the level component.} Specifically, consider the decomposition as in \eqref{eq:TVP-SIRD-decompose}, where the level component evolves according to the following factor structure
\vspace{-0.2cm}
\begin{equation}
\label{eq:TVP-SIRD-level_Factor}
    \begin{array}{rcl}
\theta_{i,l,t}   &=&  \tau_{i,l}\theta_{l,t} + \hat{\theta}_{i,l,t} \\
\theta_{l,t} &=&  \theta_{l,t-1}   + \alpha_{\theta_l} s_{\theta_l, t-1} \\
\hat{\theta}_{i,l,t} &=& \hat{\theta}_{i,l,t-1} + \alpha_{\hat{\theta}_{i,l}}s_{\hat{\theta}_{i,l}, t-1}
   \end{array}
\end{equation}
for parameter $\theta_t = \tilde{\beta}_t, \tilde{\gamma}_t$ and $\tilde{\nu}_t$, and $\theta_{l,t}$ is the common level component, respectively. Here, the key difference between the factor model and a country-specific model is that common level factor $\theta_{l,t}$ is loaded by all country-specific information with the coefficient $\tau_{i,l}$ for country $i$, \footnote{In this extension, we opt for a factor structure in the parameters similar to seasonality modeling. Alternatively, we could also proceed with a factor representation of the data, using principal components, for example, and a SIRD model corresponding to each component, similar to factor GARCH models; see \cite{FactorGARCH-09} for such a strategy.} and thus, the corresponding score function becomes
	\begin{equation}
	\label{eq:ScoreFunctions_Factor}
        \begin{array}{rcl}
    s_{\tilde{\beta},t} &=& \frac{\nabla_{1,\tilde{\beta},t} + \dots +  \nabla_{i,\tilde{\beta},t} + \dots +  \nabla_{K,\tilde{\beta},t}}{ S_{\tilde{\beta},t} } \\
    s_{\tilde{\gamma},t} &=& \frac{\nabla_{1,\tilde{\gamma},t} + \dots +  \nabla_{i,\tilde{\gamma},t} + \dots +  \nabla_{K,\tilde{\gamma},t}}{ S_{\tilde{\gamma},t} } \\
    s_{\tilde{\nu},t} &=& \frac{\nabla_{1,\tilde{\nu},t} + \dots +  \nabla_{i,\tilde{\nu},t} + \dots +  \nabla_{K,\tilde{\nu},t}}{ S_{\tilde{\nu},t} } 
    	\end{array}   
     \end{equation}  \vspace{-0.2cm}
where  \vspace{-0.4cm}
	\begin{equation}
	\label{eq:nablas_Factor}
        \begin{array}{rcl}
\nabla_{i,\tilde{\beta},t} &=& \left( \Delta C_{i,t}-\lambda_{i,\beta,t} \right)\tau_{1,i} \\
\nabla_{i,\tilde{\gamma},t} &=& \left( \Delta Rc_{i,t}-\lambda_{i,\gamma,t} \right)(1-\gamma_{i,t})\tau_{2,i} \\
\nabla_{i,\tilde{\nu},t} &=& \left( \Delta D_{i,t}-\lambda_{i,\nu,t} \right)(1-\nu_{i,t})\tau_{3,i}
    	\end{array}   
     \end{equation} 
and     
	\begin{equation}
	\label{eq:nablas_Factor}
        \begin{array}{rcl}
    S_{\tilde{\beta},t} &=&   \lambda_{1,\beta,t}\tau_{1,1}^2 +  \dots +  \lambda_{K,\beta,t}\tau_{1,K}^2 \\
S_{\tilde{\gamma},t}&=& \lambda_{1,\gamma,t}(1-\gamma_{1,t})^2\tau_{2,1}^2 +  \dots +  \lambda_{K,\gamma,t}(1-\gamma_{K,t})^2\tau_{2,K}^2  \\
 S_{\tilde{\nu},t} &=& \lambda_{1,\nu,t}(1-\nu_{1,t})^2\tau_{3,1}^2 +  \dots +  \lambda_{K,\nu,t}(1-\nu_{K,t})^2\tau_{3,K}^2. 
    	\end{array}   
     \end{equation}
The score functions of the idiosyncratic parts are the same as in the TVP-SIRD model. We provide details on these derivations in Section D.2 of the supplementary material. We denote this model as the factor TVP-SIRD model. In the following application, we only consider a factor structure in the infection rate, $\beta_t$, keeping the remaining parameters as wholly idiosyncratic as before. However, the extension of the factor structure to the remaining parameters is similar. Still, the parameters in their current form are not identified, as none of the components are observed. To identify the factors and idiosyncratic components separately, we set $\tau_1$ as one for the first country and fix the initial condition for the common factor, which enables the identification of the location of the factor and idiosyncratic components separately. We consider four countries in the application: US, Germany, Italy, and Brazil. We display the evolution of the number of daily active infected cases for Germany, Italy, and Brazil in Figure~\ref{fig:Factor-AIC}, while we provide a comprehensive analysis of the US in previous sections.
\begin{center}
[Insert Figure~\ref{fig:Factor-AIC} about here]
\end{center}
Figure~\ref{fig:Factor-AIC} indicates that the pandemic's evolution in Europe, represented by Germany and Italy follows a similar trajectory. On the other hand,  Brazil's pandemic trajectory exhibits a unique pattern, counter to Germany and Italy at times. Still, the countries' patterns converge with the last wave, i.e., the Omicron wave. We display the estimates of the fixed parameters related to the common factor in Table~\ref{tab:Factor-FixedParameters}. We display the evolution of the common factor infection rate and country-specific effective reproduction rates, $eR_{t}$s, in Figure~\ref{fig:Factor-Output}. 
 \begin{center}
[Insert Table~\ref{tab:Factor-FixedParameters} and Figure~\ref{fig:Factor-Output} about here]
\end{center}
Table~\ref{tab:Factor-FixedParameters} indicates that the common factor is affected by the past score function, derived in \eqref{eq:ScoreFunctions_Factor}, considerably with a coefficient close to 0.6 leading to a time-varying pattern. However, the 95\% HPDI covers a wide range of values between 0.5 and 0.8. Factor loadings for Germany and Italy are very similar, as expected, with values around 0.8 and bear little uncertainty. The lowest loading is for Brazil with a value of 0.3 with almost 0 and 0.5 for the bounds of 95\% HPDI.     

The upper left panel of Figure~\ref{fig:Factor-Output} displays the evolution of the common factor of infection rate. Following the estimates of factor loadings, this factor is mainly influenced by the pandemic trajectory of the US, Germany, and Italy. We can observe that the common factor can nicely capture all significant waves with the relatively higher values corresponding to the initial wave at the onset of the pandemic and the recent two waves, i.e., Delta and Omicron waves. The common impact of these two variants can also be observed by the increased $eR_t$ of the US, Germany, and Italy, particularly for the Delta variant. Finally, the relatively more idiosyncratic behavior of the pandemic in Brazil can be traced by the corresponding $eR_t$ in the lower right panel of Figure~\ref{fig:Factor-Output}. For Brazil, the $eR_t$ fluctuates around one for a large part of the sample period leading to the unique pattern of active infected cases as shown in the most right panel of Figure~\ref{fig:Factor-AIC}. These results show the efficacy of the factor TVP-SIRD model in capturing both the common and idiosyncratic patterns of the Covid-19 pandemic.

\section{Conclusion}
\label{sec:Conclusion}
This paper puts forward the time-varying parameters SIRD model for timely and accurate measurement of the pandemic's current stance and accurate predictions of its future trajectory. Our modeling framework falls into the class of 'generalized autoregressive score models'. These models involve parameters evolving deterministically according to an autoregressive process in the direction implied by the score function. Therefore the resulting approach permits a flexible yet parsimonious and statistically coherent framework to operate efficiently in scarce data environments. We demonstrate the proposed model's potential using daily and weekly US data using full sample estimation and out-of-sample forecasting using a recursive real-time prediction exercise. 

Our results show that the proposed framework can nicely track the stance of the pandemic. Our findings suggest that there is considerable fluctuation in the rate of infection and death rates. We further extend the model to include the infected individuals who do not show symptoms and are therefore not diagnosed. We show that this sample selection might have a sizable impact on the estimated reproduction rate. We extend the model framework in various directions, including a mixed-frequency setup blending daily and weekly Covid-19 pandemic-related critical data and a factor model setup by blending datasets from various countries. Results indicate the potential of our flexible model structure.         

\newpage

\singlespacing
	
\bibliographystyle{jae}
\bibliography{ref}

\newpage

\section*{Tables and Figures}

\vspace{-0.8cm}

	\begin{table}[H]
	\singlespacing
	\footnotesize
	\setlength{\tabcolsep}{12pt}
		\centering
\begin{threeparttable}		
		\caption{Estimation results of the SIRD model with fixed parameters and the TVP-SIRD model}
				\label{tab:FP-SIR-Parameters-TVP_alpha}
			\begin{tabular}{lcccc|ccc}\hline \hline
		&	\multicolumn{4}{l}{\underline{Panel A: Fixed parameters SIRD model}} &	\multicolumn{3}{l}{\underline{Panel B: TVP-SIRD model}}  \\\addlinespace
          & $\beta_l$ &  $\gamma_l~(\times10^{-1}) $  & $\nu_l~(\times10^{-2})$ & $R_0$  & $\alpha_{\beta_t}$ & $\alpha_{\gamma_t}$ & $\alpha_{\nu_t}$ \\ \hline
 Median   &  0.0122   & 0.0746  &  0.0133 &  1.6392 & 0.4822 & 0.6334 & 0.3514 \\    
2.5\% per.  &  0.0120   & 0.0744  &  0.0131 &  1.6155 & 0.4553 & 0.6253 & 0.3375 \\	   
97.5\% per. &  0.0124   & 0.0747  &  0.0134 &  1.6596 & 0.5104 & 0.6421 & 0.3682 \\	\hline \hline 
		\end{tabular}
   \begin{tablenotes}
         \item \footnotesize {\it Note:} The table displays the estimation results of the model in \eqref{eq:FPSIR}. We display the posterior median and the 2.5\% and 97.5\% percentiles of the posterior distributions of the corresponding parameter shown in the first row. 
      \end{tablenotes}
\end{threeparttable}  		
	\end{table} 
	
	\vspace{-0.5cm}

	\begin{table}[H]
	\setlength{\tabcolsep}{9pt}	
	\singlespacing
		\centering
\begin{threeparttable}			
		\caption{Relative RMSFEs of the competing models relative to the TVP-SIRD model - Daily confirmed cases}
			\label{tab:OOS-C}
		%\begin{turn}{270}	
			\begin{tabular}{lcccccc}  
   \hline \hline
			 & RW$-30$ & RW$-45$ & RW$-60$ & KF & TVP-SIRD-$\beta$   \\
    \hline
            $h=~1$   & 2.111 &  2.414  &  2.826                  &  1.443                     &  1.251 \\
            $h=~5$   &  1.747 &  1.959  &  2.162                  &  1.206                     &  1.183 \\
            $h=10$   &  1.183 &  1.550  &  1.562                  &  1.197                     &  1.088 \\
            $h=15$   &  1.325 &  1.317  & \textbf{1.142} & \textbf{0.971} &  1.027 \\
            $h=20$   & \textbf{1.054} & \textbf{1.024} & \textbf{ 0.798} & \textbf{ 0.865} &  \textbf{ 0.967} \\
            $h=25$   & \textbf{1.013} & \textbf{0.960} & \textbf{0.857} & \textbf{0.829} &  \textbf{1.002} \\
            $h=30$   & \textbf{0.949} & \textbf{0.862} & \textbf{0.662} & \textbf{0.774} & \textbf{0.973} \\
			\hline \hline
		\end{tabular}
   \begin{tablenotes}
         \item \footnotesize {\it Note:} The table displays the Root Mean Squared Forecast Errors (RMSFE) of the competing models in predicting the daily confirmed cases relative to the TVP-SIRD model introduced in \eqref{eq:TVP-SIRD}. RW-M stands for Rolling Window with $M$ observations as the sample size for $M=30,45,60$. KF stands for the time-varying parameter version of the SIRD model using a state space model framework with Normal error terms. TVP-SIRD-$\beta$ denotes the restricted version of the TVP-SIRD model, where we allow for time variation only in the infection rate, $\beta$. Statistical significance of relative Root Mean Squared Forecast Errors (RMSFE) is tested using the Diebold-Mariano (DM) test using the measures of squared forecast error contributions together with the HAC covariance matrix and a finite sample correction, \cite{harvey1997testing}. The bold values are statistically {\bf INsignificant} at the conventional significance level of 5\%.  
\end{tablenotes}
\end{threeparttable}  		
	\end{table} 
 
\vspace{-0.5cm}

	\begin{table}[H]
	\setlength{\tabcolsep}{9pt}	
	\singlespacing
		\centering
\begin{threeparttable}			
		\caption{Relative RMSFEs of the competing models relative to the TVP-SIRD model - Daily death cases}
			\label{tab:OOS-D}
		%\begin{turn}{270}	
			\begin{tabular}{l cccccc}  \hline \hline
			 & RW$-30$ & RW$-45$ & RW$-60$ & KF & TVP-SIRD-$\beta$   \\  \hline 
    $h=~1$ & \textbf{ 0.974} &\textbf{ 1.051} &                  1.310 &\textbf{ 0.776} & 3.055 \\
    $h=~5$ & \textbf{ 0.949} &\textbf{ 0.995} &                  1.288 &                  1.285 & 2.852 \\
    $h=10$ & \textbf{ 1.183} &\textbf{ 1.060} &                  1.413 &                  1.332 & 2.799 \\
    $h=15$ &                   1.147 &                  1.105 &                  1.336 &                  1.175 & 2.642 \\
    $h=20$ &                  1.130 &                  1.170 &                  1.326 &                  1.186 & 2.566 \\
    $h=25$ &                  1.096 &                  1.110 &                  1.214 &                  1.301 & 2.266 \\
    $h=30$ & \textbf{ 1.074} &\textbf{1.068} &\textbf{ 1.097} &                  1.294 & 2.099 \\
			\hline  \hline 
		\end{tabular}  						
   \begin{tablenotes}
         \item \footnotesize {\it Note:} The table displays the Root Mean Squared Forecast Errors (RMSFE) of the competing models in predicting the daily death cases relative to the TVP-SIRD model introduced in \eqref{eq:TVP-SIRD}. RW-M stands for Rolling Window with $M$ observations as the sample size for $M=30,45,60$. KF stands for the time-varying parameter version of the SIRD model using a state space model framework with Normal error terms. TVP-SIRD-$\beta$ denotes the restricted version of the TVP-SIRD model, where we allow for time variation only in the infection rate, $\beta$. Statistical significance of relative Root Mean Squared Forecast Errors (RMSFE) is tested using the Diebold-Mariano (DM) test using the measures of squared forecast error contributions together with the HAC covariance matrix and a finite sample correction, \cite{harvey1997testing}. The bold values are statistically {\bf INsignificant} at the conventional significance level of 5\%. 
\end{tablenotes}
\end{threeparttable}  		
	\end{table} 

\vspace{-0.5cm}

 	\begin{table}[H]
	\setlength{\tabcolsep}{9pt}	
	\singlespacing
		\centering
\begin{threeparttable}			
		\caption{Number of models in the Forecast Hub that the TVP-SIRD model outperforms}
			\label{tab:ForecastHub}
	\begin{tabular}{l cccc}  \hline \hline
			       & 1-week & 2-week & 3-week & 4-week   \\  \hline 
    Confirmed cases (19) &  11  &  5 &  3 & 2 \\
    Death cases     (28) &  19  & 14 &  9 & 5   \\     
			\hline  \hline 
		\end{tabular}
   \begin{tablenotes}
         \item \footnotesize {\it Note:} The table displays the number of the models in the Forecast Hub with more than 30 predictions that have greater RMSFE than the TVP-SIRD model. The total number of the models considered in the comparison is indicated in the left column in the parenthesis.    
\end{tablenotes}
\end{threeparttable}  		
	\end{table}

 	\begin{table}[H]
	\singlespacing
	\footnotesize
	\setlength{\tabcolsep}{12pt}
		\centering
\begin{threeparttable}		
		\caption{Estimation results of the factor TVP-SIRD model}
				\label{tab:Factor-FixedParameters}
			\begin{tabular}{lcccc}\hline \hline \addlinespace
          &$\alpha_{\beta_{l,t}}$ & $\tau_{Ger}$ & $\tau_{It}$ & $\tau_{Br}$ \\ \hline
 Median   &  0.6347   & 0.7840  &  0.8228 &  0.2982    \\    
2.5\% per.  &  0.4944  & 0.7409 &  0.7522  &  0.0722    \\	   
97.5\% per. &  0.7750   &0.8271 &  0.8934 &  0.5242   \\	\hline \hline 
		\end{tabular}
   \begin{tablenotes}
         \item \footnotesize {\it Note:} The table displays the estimation results of the model in \eqref{eq:Factor-TVP-SIRD}. We display the posterior median and the 2.5\% and 97.5\% percentiles of the posterior distributions of the corresponding parameter shown in the first row. 
      \end{tablenotes}
\end{threeparttable}  		
	\end{table}

\newpage

 \vspace{-0.3cm}
\footnotesize \noindent {\it Note:} The graphs show the evolution of the daily confirmed and death cases in the US over the sample from March 2020 until the end of March 2022.	

 \vspace{+1cm}

 \vspace{-0.3cm}
\footnotesize \noindent {\it Note:} The graphs show the evolution of the effective reproduction rate, $eR_t$, estimated using the SIRD model with fixed parameters displayed in \eqref{eq:FPSIR} in the US over the sample from March 2020 until the end of March 2022.	The 95\% (HPDI) Highest Posterior Density Intervals are computed using the posterior output.  

 \vspace{+1cm}

 \vspace{-0.3cm}
\footnotesize \noindent {\it Note:} The graphs show the evolution of the daily confirmed and death cases in the US and the fitted values using the TVP-SIRD model over the sample from March 2020 until the end of March 2022.

 \vspace{-0.5cm}
\footnotesize \noindent {\it Note:} The graphs show the evolution of the time-varying parameters, $\beta_{l,t}$, the rate of infection $\nu_{l,t}$, the death rate, and the effective reproduction rate, $eR_t$, estimated using the TVP-SIRD model introduced in \eqref{eq:TVP-SIRD} for the US. The 95\% (HPDI) Highest Posterior Density Intervals are computed using the posterior output.  

 \vspace{-0.5cm}
\footnotesize \noindent {\it Note:} The graph shows the evolution of the ratio of total infections to the number of reported infected cases estimated using the MF-TVP-SIRD model introduced in \eqref{eq:MF-TVP-SIRD} for the US. The 95\% (HPDI) Highest Posterior Density Intervals are computed using the posterior output.

  \vspace{-0.3cm}
\footnotesize \noindent {\it Note:} The graphs show the evolution of the time-varying parameters, $\nu_{l,t}$, the death rate, and $eR_{l,t}$, the effective reproduction rate, estimated using the MF-TVP-SIRD model introduced in \eqref{eq:MF-TVP-SIRD} for the US. The 95\% (HPDI) Highest Posterior Density Intervals are computed using the posterior output.

 \vspace{-0.5cm}
\footnotesize \noindent {\it Note:} The graphs show the evolution of the relative Root Mean Squared Forecast Error (rRMSFE) for the weekly 1-week ahead predictions of the ensemble model from Forecast-Hub relative to the TVP-SIRD model estimated using weekly data for the period starting from July 2020 until March 2022. The solid line shows the rRMSFEs computed using expanding window. The dashed line indicates actual realizations of weekly confirmed and death cases.

 \vspace{-0.5cm}
\footnotesize \noindent {\it Note:} The graphs show the evolution of the active infected cases in Germany, Italy, and Brazil over the sample from March 2020 until the end of March 2022. The number of recovered cases is absent in all countries in half of the sample. Therefore, we use $\gamma=0.07$, corresponding to a recovery duration of 14 days when computing the active infected cases.

 \vspace{-0.5cm}
\footnotesize \noindent {\it Note:} The graphs show the evolution of the common infection rate and country-specific $eR_t$s for Germany, Italy, and Brazil over the sample from March 2020 until the end of March 2022 using the factor TVP-SIRD model as introduced in \eqref{eq:Factor-TVP-SIRD}.

\end{document}